\documentclass[aps,prx,preprint,showkeys,nofootinbib,superscriptaddress]{revtex4-2}

\usepackage[english]{babel}
\usepackage[utf8]{inputenc}
\usepackage{amsmath}
\usepackage{xcolor}
\usepackage{graphicx}
\usepackage{caption}
\usepackage{subcaption}
\usepackage[hidelinks]{hyperref}

\begin{document}

\bibliographystyle{apsrev4-2}

\title{Machine-Learned Interatomic Potentials for Structural and Defect Properties of YBa$_2$Cu$_3$O$_{7-\delta}$}

\author{Niccolò Di Eugenio$^{\dagger,\ast}$}
\affiliation{Department of Applied Science and Technology, Politecnico di Torino, I-10129 Torino, Italy}
\affiliation{Istituto Nazionale di Fisica Nucleare, Sezione di Torino, I-10125 Torino, Italy}

\author{Ashley Dickson$^{\dagger}$}
\affiliation{School of Engineering, Lancaster University, Lancaster, LA1 4YW, United Kingdom}

\author{Flyura Djurabekova}
\affiliation{Department of Physics, University of Helsinki, P.O. Box 43, FI-00014, Finland}

\author{Francesco Laviano}
\affiliation{Department of Applied Science and Technology, Politecnico di Torino, I-10129 Torino, Italy}
\affiliation{Istituto Nazionale di Fisica Nucleare, Sezione di Torino, I-10125 Torino, Italy}

\author{Federico Ledda}
\affiliation{Department of Applied Science and Technology, Politecnico di Torino, I-10129 Torino, Italy}
\affiliation{Istituto Nazionale di Fisica Nucleare, Sezione di Torino, I-10125 Torino, Italy}
\affiliation{Department of Physics, University of Helsinki, P.O. Box 43, FI-00014, Finland}

\author{Daniele Torsello}
\affiliation{Department of Applied Science and Technology, Politecnico di Torino, I-10129 Torino, Italy}
\affiliation{Istituto Nazionale di Fisica Nucleare, Sezione di Torino, I-10125 Torino, Italy}

\author{Erik Gallo}
\affiliation{Eni S.p.A., Piazzale Enrico Mattei, 1, I-00144 Roma, Italy}

\author{Mark R. Gilbert}
\affiliation{United Kingdom Atomic Energy Authority, Culham Campus, Abingdon, OX14 3DB, United Kingdom}

\author{Duc Nguyen-Manh}
\affiliation{United Kingdom Atomic Energy Authority, Culham Campus, Abingdon, OX14 3DB, United Kingdom}

\author{Antonio Trotta}
\affiliation{Eni S.p.A., MAFE, I-30175 Venice, Italy}

\author{Samuel T. Murphy}
\affiliation{School of Engineering, Lancaster University, Lancaster, LA1 4YW, United Kingdom}

\author{Davide Gambino}
\affiliation{Department of Physics, University of Helsinki, P.O. Box 43, FI-00014, Finland}
\affiliation{Theoretical Physics Division, Department of Physics, Chemistry, and Biology (IFM), Linköping University, Linköping 58183, Sweden}

\date{\today}

\begin{abstract}
High-Temperature Superconductors (HTS) such as YBa$_2$Cu$_3$O$_{7-\delta}$ (YBCO) are essential for next-generation Tokamak fusion reactors, where Rare-Earth Barium Copper Oxides (REBCO) form the functional layers in HTS magnets. Because YBCO’s superconductivity depends strongly on oxygen stoichiometry and defect structure, atomistic simulations can provide crucial insight into radiation‐damage mechanisms and pathways to maintain material performance. 

In this work, we develop and benchmark four Machine-Learned Interatomic Potentials (MLPs) for YBCO—the Atomic Cluster Expansion (ACE), the Message-Passing Atomic Cluster Expansion (MACE),the Gaussian Approximation Potential (GAP), and the Tabulated Gaussian Approximation Potential (tabGAP)—trained on an extensive Density Functional Theory (DFT) database explicitly designed to include irradiation-damaged-like configurations. The resulting models achieve DFT-level accuracy across a wide range of atomic environments, faithfully capturing the interatomic forces relevant to radiation damage processes. 

Among the tested models, MACE delivers the highest accuracy, however at greater computational cost, while ACE and tabGAP provide an excellent balance between efficiency and fidelity. These machine-learned potentials establish a robust foundation for large-scale Molecular Dynamics (MD) simulations of radiation-induced defect evolution in complex superconducting materials.
\end{abstract}

\maketitle

\footnotetext{$\dagger$ These authors contributed equally to this work.}
\footnotetext{$\ast$ Contact author: \href{mailto:niccolo.dieugenio@polito.it}{niccolo.dieugenio@polito.it}}

\section{Introduction} \label{sec:introduction}

\begin{figure}[ht!]
    \centering
    \includegraphics[width=0.5\linewidth]{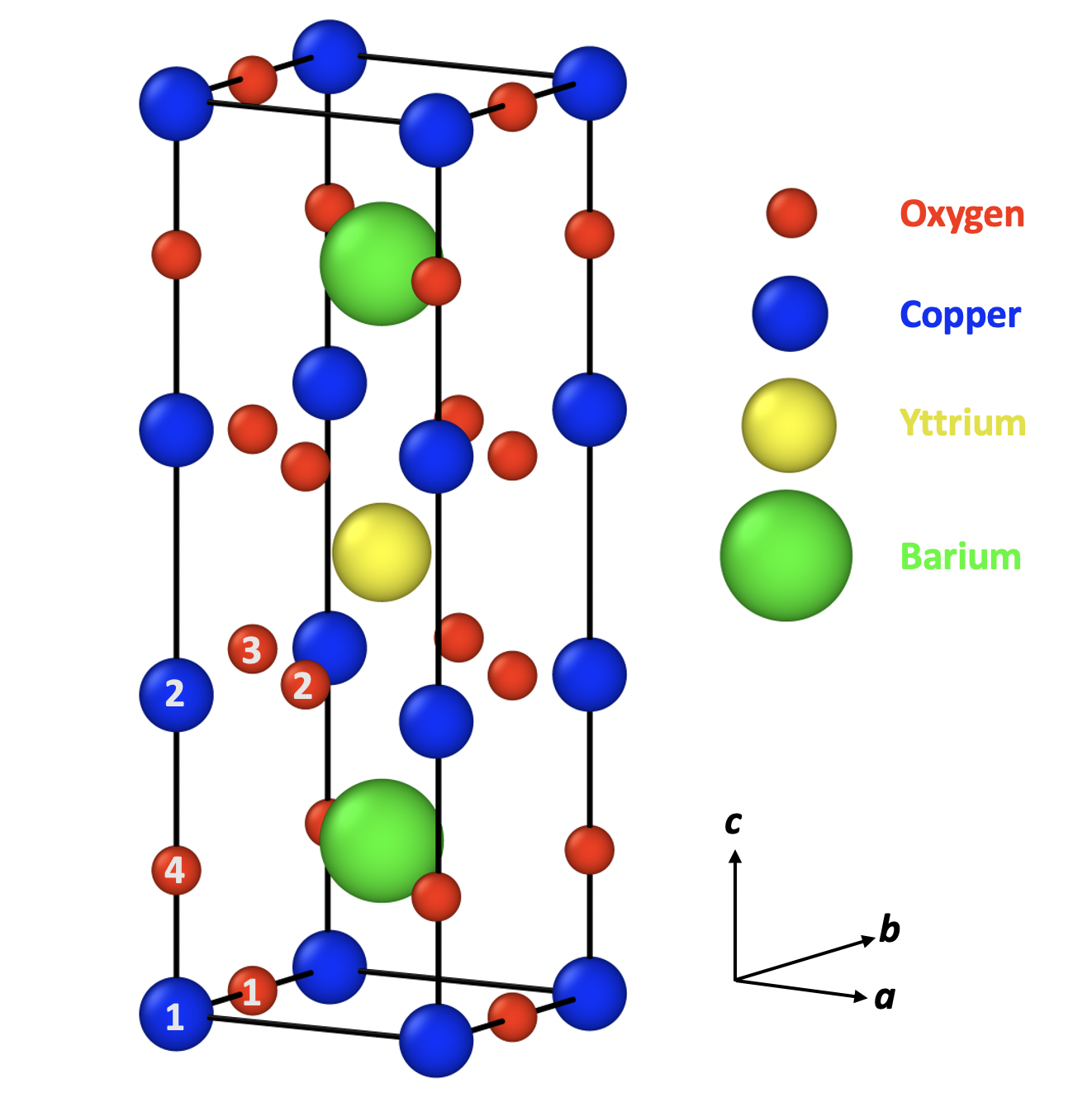}
    \caption{Structure diagram of YBa$_2$Cu$_3$O$_7$. The numbers contained within atoms are the site numbers for symmetrically distinct O/Cu atoms.}
    \label{fig:rebcostruct}
\end{figure}

High-Temperature Superconductors (HTS) are an enabling technology for magnetic plasma
confinement \cite{Mitchell2021, Kuang2018}, which is crucial for the design of compact fusion reactors, as HTS magnets are built from composite tapes, generally employing REBCOs as the functional material. REBCO compounds possess a highly anisotropic orthorhombic structure (space group \textit{Pmmm}(47), as shown in Fig.~\ref{fig:rebcostruct}) featuring Cu1 sites forming Cu–O chains and Cu2 sites in square-pyramidal coordination within Cu–O$_2$ planes.

YBa$_2$Cu$_3$O$_{7-\delta}$ (YBCO) is a compound from the REBCO group of materials, whose characteristics are strongly influenced by composition and atomic structure \cite{Johnson2008} and are subject to profound changes when crystal defects are present \cite{Nucker1988}. The oxygen sublattice stoichiometry is strongly linked to superconducting behavior \cite{fischer2018effect}, varying from an antiferromagnetic insulator at $\delta = 1$ to a superconductor for $\delta$ values below $\sim0.6$ \cite{Tallon2005,cava1987oxygen}. Given that neutron irradiation in fusion reactors induces structural damage \cite{Iliffe2021} (via collision cascades) and affects superconducting performance \cite{fischer2018effect} it is important to understand the underlying mechanisms behind this degradation. Atomistic simulations can thus provide insight into the microscopic mechanisms of radiation damage, elucidate irradiation effects \cite{Torsello2025}, and suggest strategies to maintain material performance \cite{Torsello2023}. Furthermore, the insights gained from these simulations inform engineering design decisions in fusion devices, such as optimizing shielding configurations to protect superconducting magnets \cite{Calzavara2025}.

Theoretical attempts to model the thermodynamic properties of these complex materials have been made using Density Functional Theory (DFT) \cite{murphy2020point, liu2023first, nicholls2022understanding, Dickson2026}. However, these DFT simulations are inherently constrained to short time and length scales, rendering them unsuitable for the direct modeling of high-energy, non-equilibrium phenomena associated with radiation-induced damage. Machine-Learned Potentials (MLPs) represent a data-driven approach to interpolate between high-fidelity data (typically DFT), allowing the construction of highly accurate interatomic potentials that can reproduce DFT-level accuracy under any atomic environment, including those generated in radiation conditions \cite{byggmastar2022multiscale, zongo2024unified}. As such, MLPs open the possibility of modeling defect generation, cascade evolution, and energy dissipation processes in HTS materials at a level of precision previously inaccessible to classical interatomic potentials.

Only a few examples of MLPs trained on materials with comparable structural and chemical complexity have been reported \cite{zhao2023complex,zongo2024unified,attarian2024studies}, primarily due to the challenge of sampling such an extensive compositional space. To the best of our knowledge, YBCO represents one of the most chemically and structurally complex materials for which a hand-crafted machine-learning potential has been successfully developed. This complexity necessitates the development of a bespoke database and requires careful parametrization \cite{Bochkarev2022}. Whilst powerful foundation models have recently been developed for chemically and structurally complex materials - such as M3GNet15 \cite{Chen2022}, CHGNet16 \cite{Deng2023}, GNoME17 \cite{Merchant2023}, and MACE-MP-0 \cite{batatia2025foundation} - their accuracy in specialized applications still falls short of that achievable with a carefully constructed, system-specific potential \cite{Radova2025, Marchand2025}. Fine-tuning such foundation models \cite{radova2025fine} can improve performance, but often at the cost of reduced transferability, particularly in highly non-equilibrium environments such as those encountered under irradiation—conditions where a dedicated potential remains essential. There is indeed still a need for a potential that is flexible enough to accurately describe both stoichiometric and non-stoichiometric YBCO, while also capturing the behavior of highly disordered collision cascades. To this goal, we adopt a hand-made approach, in which we create a database purposely conceived to target specific physical properties in YBCO, as well as more general predictions.

In this work, we develop and compare four types of MLP architectures—the Atomic Cluster Expansion (ACE) \cite{Drautz2019, Drautz2020}, the Message-Passing Atomic Cluster Expansion (MACE) \cite{Batatia2022}, the Gaussian Approximation Potential (GAP) \cite{deringer2021gaussian}, and the Tabulated Gaussian Approximation Potential (tabGAP) \cite{Byggmstar2022}—with the goal of developing reliable interatomic potentials for predicting the effects of radiation damage in YBCO. The potentials are fit to a purpose-built DFT database, designed to include configurations representative of both equilibrium and highly non-equilibrium (irradiation-damaged-like) environments. This ensures that the learned models accurately reproduce interatomic forces over a broad energy landscape, including those associated with defect production and radiation-induced disorder.

The developed potentials were rigorously validated against the equation of state under various pressures, elastic constants, bulk modulus, defect formation energies, lattice parameters as a function of oxygen concentration, and activation barriers from Nudged Elastic Band (NEB) calculations \cite{Henkelman2000}. In addition, the models correctly capture the orthorhombic–tetragonal phase transition, which occurs at elevated temperatures, or under conditions of oxygen depletion. This transition has been observed experimentally \cite{Jorgensen1987} and reproduced only at high temperatures by previous interatomic potentials for YBCO \cite{Chaplot1989, Chaplot1988, gray2022molecular}, as well as by the first MLP for YBCO \cite{Gambino2025}. The latter was developed with the specific focus on capturing the phase transition. In contrast to the models presented in this work, existing potentials are unable to treat sub-stoichiometric YBCO and consequently do not reproduce the phase transition associated with oxygen depletion.

We demonstrate that MLPs developed in this working surpass existing interatomic potentials - developed by Gray et al. \cite{gray2022molecular}, Baetzold \cite{baetzold1988atomistic}, and Chaplot \cite{Chaplot1989} - in both computational efficiency and accuracy in describing thermodynamic and defect properties, while also capturing the orthorhombic–tetragonal phase transition with improved stability. 

More importantly, the developed MLPs exhibit excellent agreement with DFT-calculated forces across a wide range of atomic configurations, including those relevant to displacement cascades and radiation-induced defect formation. This makes them ideally suited for future large-scale simulations of irradiation effects in HTS, providing a faithful description of interatomic forces and energies under extreme conditions, such as those observed in harsh environments.

Finally, these MLPs enable the investigation of defect dynamics and radiation damage processes that are otherwise computationally prohibitive to study with DFT. They accurately describe complex, highly-defected materials and capture changes in their physical properties under irradiation with DFT-level fidelity. Given that all MLPs were trained on the same database, our results enable a direct comparison of their accuracy, transferability, and stability. This demonstrates the potential of data-driven models to serve as robust tools for radiation-damage studies in HTS materials, including analyses of defect distributions and morphologies via Molecular Dynamics (MD) simulations.

\section{Methods} \label{sec:methods}

\subsection{Potentials Formalisms}

\subsubsection{Atomic Cluster Expansion (ACE)}

Traditional cluster expansions express the total energy of an atomic configuration as a many-body series over increasing cluster sizes \cite{Drautz2004},
\begin{equation}
E = V_0 + \sum_i V^{(1)}(i) + \frac{1}{2}\sum_{ij} V^{(2)}(i,j) + \frac{1}{3!}\sum_{ijk} V^{(3)}(i,j,k) + \cdots .
\end{equation}
where $V^{(K)}$ represents an effective $K$-body interaction. While this formal expansion is exact, its direct evaluation becomes challenging since the computational cost grows exponentially with the number of neighbors and interaction order.

ACE \cite{Drautz2019} reformulates this idea into a numerically efficient and systematically convergent framework for representing local atomic environments. Each atomic energy is expanded in a complete orthogonal basis of functions that depend on the relative positions of neighboring atoms, combining radial basis functions $R_{nl}(r)$ with spherical harmonics $Y_{lm}(\hat{r})$. The resulting basis functions are symmetrized over all neighbors, ensuring invariance with respect to translation, rotation, and permutation of identical atoms. The atomic energy of atom $i$ can then be expressed as
\begin{equation}
E_i = \sum_n c^{(1)}_n B^{(1)}_{in} + \sum_{n_1 n_2 l} c^{(2)}_{n_1 n_2 l} B^{(2)}_{i n_1 n_2 l} + \sum_{n_1 n_2 n_3 l_1 l_2 l_3} c^{(3)}_{n_1 n_2 n_3 l_1 l_2 l_3} B^{(3)}_{i n_1 n_2 n_3 l_1 l_2 l_3} + \cdots ,
\end{equation}
where $B^{(K)}$ are invariant many-body basis functions of body order $K$. The linear combination coefficients $c^{(K)}$ are obtained by fitting to quantum-mechanical reference data.

In the present work, the ACE basis is constructed using simplified Bessel radial functions together with a shifted–scaled Finnis--Sinclair–type embedding to define the density channels in the polynomial expansion.

A key advantage of ACE is that the cost of evaluating the basis scales linearly with the number of neighbors, independent of the expansion order. This allows inclusion of higher-body terms that capture complex angular correlations at modest computational expense. Nonlinear extensions of ACE can also recover the accuracy of modern machine-learning force fields while maintaining a physically interpretable form. In this way, ACE serves as a bridge between traditional analytic potentials and high-accuracy machine-learning models.

\subsubsection{Message-Passing Atomic Cluster Expansion (MACE)}

MACE \cite{Batatia2022} extends ACE by combining the many-body hierarchy of local descriptors with equivariant Graph Neural Networks (GNNs) \cite{Gilmer2017, Bronstein2021}. MACE is theoretically based on the Multi-ACE framework \cite{Batatia2025}, where messages are constructed as higher-order tensor products of atomic features, capturing $(\nu + 1)$-body correlations while maintaining exact $E(3)$-equivariance, i.e.\ correct transformation under translation, rotation, and reflection \cite{Batatia2025}. This formulation yields high expressive power with only one or two many-body equivariant layers.

The model maps atomic positions and chemical elements to potential energy by decomposing the total energy into atomic site contributions. Atoms within a cutoff radius $r_\mathrm{cut}$ define local neighborhoods,
\[
\mathcal{N}(i)=\{\, j \mid |r_{ij}| \le r_\mathrm{cut} \,\}.
\]
Node features $h_{i,kLM}^{(t)}$ are initialized from learnable species embeddings using invariant $(L=0)$ components,
\begin{equation}
h_{i,k00}^{(0)} = \sum_z W_k^z \, \delta_{z z_i},
\end{equation}
and are expanded in a spherical-harmonic basis to ensure equivariance.

Neighbor features and displacement vectors form the equivariant one-particle basis $A_i^{(t)}$, obtained by symmetrizing over neighbors to enforce permutation invariance. Higher-order equivariant features $B_i^{(t),\nu}$ are then constructed as tensor products of the $A_i^{(t)}$ features up to body order $\nu = 3$ (four-body terms).

A learnable linear combination of these many-body features defines the atomic messages:
\begin{equation}
m_{i,kLM}^{(t)} =
\sum_{\nu} 
\sum_{\eta_\nu}
W_{z_i,kL}^{(t,\nu,\eta_\nu)}
B_{i,\eta_\nu kLM}^{(t),\nu}.
\end{equation}
The learnable weights $W_{z_i,kL}^{(t,\nu,\eta_\nu)}$ correspond to trainable parameters that tie together the atomic type $z_i$, channel index $k$, angular momentum $L$, message-passing layer $t$, body-order index $\nu$, and feature-group index $\eta_\nu$.

Two message-passing layers yield an effective receptive field of $2r_\mathrm{cut}$. Atomic site energies are obtained by applying invariant readout functions to the updated node features:
\begin{equation}
E_i = \sum_{t=1}^{2} \mathcal{R}^{(t)}\!\left(h_i^{(t)}\right),
\end{equation}
where $\mathcal{R}^{(1)}$ is linear and $\mathcal{R}^{(2)}$ is implemented as a single-layer MultiLayer Perceptron. Forces and stresses follow from analytical derivatives of the total energy. By integrating high-body-order equivariant message passing with site-energy decomposition, MACE achieves excellent accuracy and transferability across molecular systems.

\subsubsection{Gaussian Approximation Potential (GAP)}

The GAP framework represents the total energy as a sum of contributions from user-defined descriptors \cite{deringer2021gaussian}. In this work, we use two-body, three-body, and Embedded Atom Method (EAM) terms \cite{byggmastar2022multiscale}:
\begin{equation}
E = \sum_{p}\epsilon_p^{(2)} + \sum_{t}\epsilon_t^{(3)} + \sum_{m}\epsilon_m^{(eam)} .
\end{equation}

Since DFT provides only total energies, GAP models the covariance between configurations $N$ and $M$ as
\begin{equation}
\langle E_N E_M \rangle = \sigma_w^2 \sum_{i\in N}\sum_{j\in M} C(\mathbf{d}_i,\mathbf{d}_j) ,
\end{equation}
where $\mathbf{d}_i$ encodes the local environment of atom $i$, and $C$ is a kernel (typically squared exponential) measuring environment similarity. Treating site energies $\epsilon(\mathbf{d})$ as samples from a Gaussian Process leads to the standard Gaussian Process Regression (GPR) prediction for a new configuration,
\begin{equation}
\bar{y} = \mathbf{k}^T \mathbf{C}^{-1}\mathbf{t} ,
\label{GPR}
\end{equation}
where $\mathbf{C}$ is the covariance matrix ($\mathbf{C}=\langle \mathbf{t}\mathbf{t}^T\rangle$) of total energy observations from the training set ($\mathbf{t}$). $\bar{y}$ is then the mean predicted value for the total energy. This is calculated using $\mathbf{k}^T$: the covariance vector of function values ($\langle y\mathbf{t} \rangle$, where $y$ is the scalar function value at the new test input). To reduce computational cost, GAP employs sparsification over a representative subset of atomic environments \cite{deringer2021gaussian}.

\subsubsection{Tabulated Gaussian Approximation Potential (tabGAP)}

Direct evaluation of Eq.~(\ref{GPR}) is computationally demanding, limiting the speed of GAP models. A key strategy to overcome this is descriptor tabulation, yielding the so-called tabGAP approach. First introduced by Glielmo \textit{et al.} \cite{glielmo2018efficient} and extended by Byggm{\"a}star \textit{et al.} \cite{byggmastar2021modeling}, tabGAP replaces kernel evaluations with spline interpolations on low-dimensional descriptor grids. Pairwise interactions are tabulated as one-dimensional functions of interatomic distance, while three-body terms are tabulated on three-dimensional grids of $(r_{ij}, r_{ik}, \theta_{ijk})$. The total energy in tabGAP then takes the form:
\begin{equation}
E^{\mathrm{tabGAP}}_{\mathrm{tot}} = 
\sum_{i<j}^N S_{ij}^{1D}(r_{ij}) +
\sum_{i,j<k}^N S^{3D}_{ijk}(r_{ij},r_{ik}, \cos{\theta_{ijk}}) .
\end{equation}
where $S^{1D}$ and $S^{3D}$ are 1D and 3D cubic spline interpolations. The repulsive and GAP pair potentials are merged into a single 1D spline for each element pair. Depending on the choice of descriptors and cutoffs, tabGAP can accelerate evaluations by several orders of magnitude compared to the standard GAP formalism.

\subsection{Density Functional Theory Calculations} \label{ssec:DFT}

DFT calculations were carried out with the quantum chemistry and solid state physics program package CP2K \cite{hutter2014cp2k}, which employs a mixed Gaussian Plane Waves approach (GPW), implemented in QUICKSTEP \cite{VandeVondele2005}. The GPW method solves the DFT Kohn-Sham equations efficiently, using Gaussians as a basis set and plane waves as an auxiliary basis. A plane-wave cutoff of 400 Ry was found to ensure energy convergence within 1 meV for all structures. Exchange–correlation effects were treated within the Generalised Gradient Approximation (GGA) using the Perdew–Burke–Ernzerhof (PBE) functional \cite{perdew1996generalized}.

Atomic orbitals were expanded in a double-zeta valence plus polarisation (DZVP) basis set. Core–valence interactions were modeled with dual-space, norm-conserving, separable, PBE-optimised Goedecker–Teter–Hutter (GTH) pseudopotentials \cite{krack2005pseudopotentials}. All simulations were performed at the $\Gamma$ point with periodic boundary conditions applied in all three cartesian directions. To generate the training data, the dimensions of the simulation cells were varied according to energy convergence for each material (see Sec.~\ref{database}). To demonstrate the strong agreement of these simulations with experiments, the relaxed lattice parameters of each training structure were calculated  (see Supplemental Material at \cite{SuppMat}). Note that the DFT training data from Gambino \textit{et al.}  \cite{Gambino2025} was generated using VASP with the Projector Augmented Wave method and the PBE exchange–correlation functional. As such, our training data could not be readily combined. 

Defect and equation of state calculations utilised a 6$\times$6$\times$2 supercell of YBCO, with the exception of the NEB calculations, which were performed in a 4$\times$4$\times$2 cell. DFT cell relaxations as a function of oxygen content in YBCO were performed in a 4$\times$4$\times$2 cell, which proved to be sufficient for good agreement with experimental results. Thermal expansion simulations were performed in a 6$\times$6$\times$2 cell. For each temperature, the system was first equilibrated for 2 ps in an NPT ensemble, followed by a 1 ps run in the same ensemble, employing a timestep of 1 fs.

\subsection{Machine Learned Potential Calculations}

All relaxations and minimization with the MLPs and the Gray \cite{gray2022molecular} potential were performed in LAMMPS \cite{Thompson2022}. All simulations for MACE, ACE, GAP and tabGAP have employed, respectively, the \verb|symmetrix| \cite{Kovacs2025, batatia2025foundation}, \verb|pace| \cite{Drautz2019, Lysogorskiy2021,Lysogorskiy2023}, \verb|quip| \cite{Bartk2010, BartkPrtay2010} and \verb|tabgap| \cite{Byggmstar2022, byggmastar2021modeling} LAMMPS \verb|pair_style|.

All defect energies, NEB, equation of state, thermal expansion and stress simulations were performed in a 6$\times$6$\times$2 supercell. The measurement of lattice parameters as a function of oxygen concentration was performed in the same cell (4$\times$4$\times$2) as DFT so the results are directly comparable, as different distributions of oxygen defects may have differing effects on lattice parameters. Thermal expansion simulations were performed in a 6$\times$6$\times$2 cell, with an equilibration time of 2 ps in an NPT ensemble. In the case of GAP and tabGAP, simulations were then allowed to run for 1 ns with a 1 fs timestep, and the lattice parameters were averaged over the region where the $a$ and $b$ lattice parameters remained stable. With MACE and ACE, simulations ran for 1 ns at T = [100,200,300,400,500,600,700,800,900] K, for 3 ns at T = [1000,1400] K and for 4 ns at T = [1100,1200,1300] K, in order to better sample the transition region and reduce noise.

\subsection{Defect Formalism}

Defect formation energies were computed as follows: first, a defect-free supercell was fully relaxed, allowing both atomic positions and cell parameters to vary. Next, a specific defect (vacancy, antisite, or interstitial) was introduced into the relaxed structure, after which the system was re-relaxed with the cell shape and volume fixed, allowing only atomic positions to relax. Finally, the defect formation energies were computed following the methodology described in \cite{murphy2020point}:
\begin{equation}
\Delta E_f = E_{\text{defect}}^{T} - E_{\text{perf}}^{T} + \sum_{\alpha} n_{\alpha} \mu_{\alpha} ,
\label{eq:defect_energy}
\end{equation}
where $E_{\text{defect}}^{T}$ and $E_{\text{perf}}^{T}$ are the total DFT energies of the system with and without the defect, respectively. The term $n_{\alpha}$ represents the number of atoms of species $\alpha$ that are added or removed to create the defect, and $\mu_{\alpha}$ is the corresponding chemical potential of species $\alpha$ (calculated from the pure elements). It should be noted that the determination of the oxygen chemical potential in DFT should include spin polarization. However, since our training database comprises only non–spin-polarized calculations, introducing spin polarization would result in an inconsistent reference energy. Therefore, for the purposes of validation (assessing how well the MLPs reproduce the training data) we use non–spin-polarized DFT calculations for O$_{2}$. Accordingly, all DFT chemical potentials reported herein are taken from non–spin-polarized calculations. Moreover, spin effects have been shown to have negligible impact on similar oxide systems \cite{Lee2025}, supporting the use of non–spin-polarized configurations here.

\subsection{Database Construction}\label{database}

We constructed a diverse training database to capture the complexity of YBCO and its subsystems. The main focus was on YBa$_2$Cu$_3$O$_6$ (YBCO$_6$) and YBa$_2$Cu$_3$O$_7$ (YBCO$_7$), enabling the model to describe sub-stoichiometric phases. Pure elements (Y, Ba, Cu, O) and selected binary oxides (Y$_2$O$_3$, BaO, Cu$_2$O) were added to provide transferability across different chemical environments. CuO was excluded due to the lack of spin treatment in our DFT setup.

We performed supercell convergence at the $\Gamma$-point (up to $\sim$1000 atoms, 1 meV energy tolerance), calculating the energy per atom and stopping when the change in energy per atom was less than \(10^{-3}\,\text{eV}\). This procedure saves unnecessarily long computations for materials that require less dense sampling in reciprocal space (i.e., a smaller cell). This arises due to differences in electronic occupation for different materials. For instance, metals exhibit sharp discontinuities in electronic occupation at the Fermi level due to the absence of a band gap, and therefore require denser reciprocal-space sampling than insulators. The final cells were then relaxed at zero pressure. Three classes of perturbations were then systematically applied:

\begin{itemize}
\item Volume expansion/compression: –10\% to +10\% in 2.5\% steps, then up to +200\% in 10\% steps (30 volumes per structure).

\item Strain and angular distortions: randomized strain (–5\% to +5\%) with normally distributed angle variations, where $2\sigma\approx 5^\circ$ (5 strains per each volume).

\item Thermal displacements: random perturbations constrained by the Lindemann criterion \cite{Lindemann1910}, which is observed by fixing a max value of 10\% of the unit cell lattice parameters (5 thermal displacement per each strain).
\end{itemize}

Each relaxed cell generated 750 configurations (30 volume $\times$ 5 strains $\times$ 5 thermal perturbations), for which single-point DFT energies, forces, and stresses were computed. Then, defects in YBCO$_7$ were extensively sampled. Seven known interstitial sites \cite{murphy2020point} were tested for all four elements (28 systems), each with small volume variations (0 to ±5\%), three random strains, and four thermal perturbations. All possible vacancies and eight cation antisite configurations were treated in the same way. Amorphous cells were generated in OVITO \cite{stukowski2009visualization} via random displacements to represent cascade damage. To improve high-temperature transferability, active learning cycles were performed: MD trajectories driven by preliminary MLPs were sampled, re-evaluated at the DFT level, and then reintegrated into the database. All the initial crystal structures were downloaded from the Materials Project database \cite{Jain2013}.

\begin{figure*}[ht!]
    \centering
    \includegraphics[width=1\linewidth]{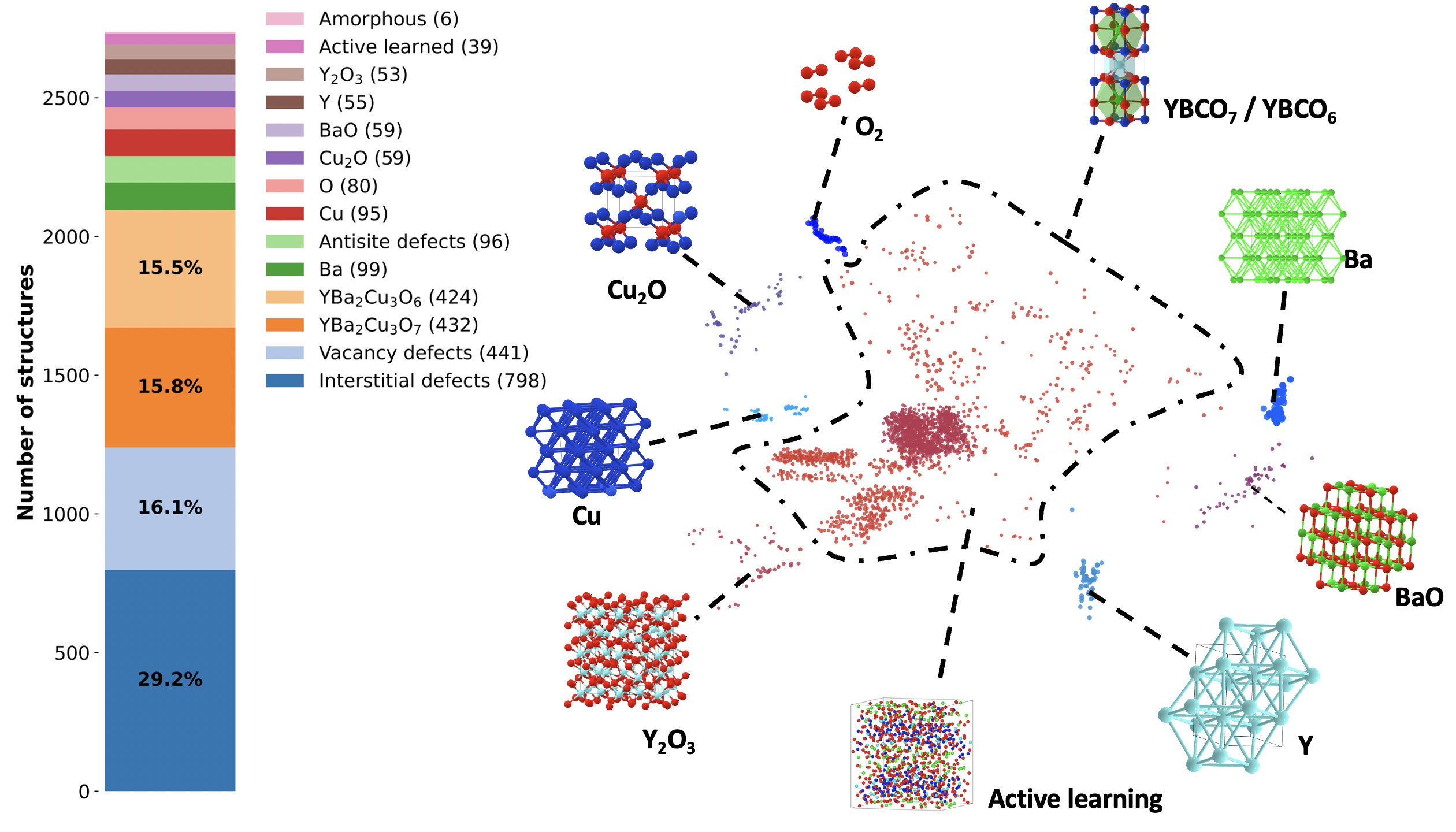}
    \caption{Overview of the structural diversity within the dataset. All structures' images are downloaded from the Materials Project \cite{Jain2013} database, apart from the active learned configuration. Each point corresponds to a structure represented by its SOAP descriptor, with pairwise similarities computed via the average kernel implemented in DScribe \cite{himanen2020dscribe}. The resulting similarity matrix was projected onto two dimensions using t-distributed stochastic neighbor embedding (t-SNE) \cite{soni2020visualizing, pedregosa2011scikit}. Three principal high-density “islands” are observed within the YBCO region (encircled by a dotted line): the left cluster corresponds to low-energy YBCO$_7$ configurations, the lower cluster to low-energy YBCO$_6$ structures, and the upper cluster to defect-containing configurations. Highly strained and distorted structures extend toward the upper-left region. Distinct structural classes are colour-coded for clarity. The accompanying composition chart shows the relative abundance of each structure type, with the number of configurations indicated in brackets.}
    \label{dataset}
\end{figure*}

Finally, highly similar configurations were filtered out of the dataset. This was crucial to reduce training time for the potentials, as the training set initially comprised of $\sim$10,000 DFT structures. First, configurations with forces $>$ 80~eV/\AA{} were discarded, as tests showed that including larger-force outliers did not improve model accuracy.
For YBCO$_7$ and YBCO$_6$, only volumes $\leq$ 100\% were retained (as tests indicated these configurations made no appreciable difference); for elemental and binary oxides, only volumes in the range –10\% to +10\% were used. Thermal perturbation sets were further reduced using farthest-point sampling on Smooth Overlap of Atomic Positions (SOAP \cite{bartok2013representing}) principal components to maximize structural dissimilarity. This was implemented by taking the mean SOAP vector of each configuration (with DScribe \cite{himanen2020dscribe}), and utilising scikit-learn \cite{pedregosa2011scikit} dimensionality reduction to output two principal components. This was carried out in a manner similar to Shenoy \textit{et al.} \cite{shenoy2024collinear}. A visual representation of the dataset is shown in Fig.~\ref{dataset}. The final training set consisted of 2712 structures, with 7,174,977 force components.

Employing the database described above, four different MLP models have been trained with MACE, ACE, GAP and tabGAP. From this point on, we will refer to the trained models as YBCO\_MACE, YBCO\_ACE, YBCO\_GAP and YBCO\_tabGAP. Details regarding hyperparameter selection for each model are found in Supplemental Material at \cite{SuppMat}.

\section{Results and Discussion}

\subsection{Initial Validation}

\begin{figure*}[ht!]
    \centering
    \includegraphics[width=\textwidth]{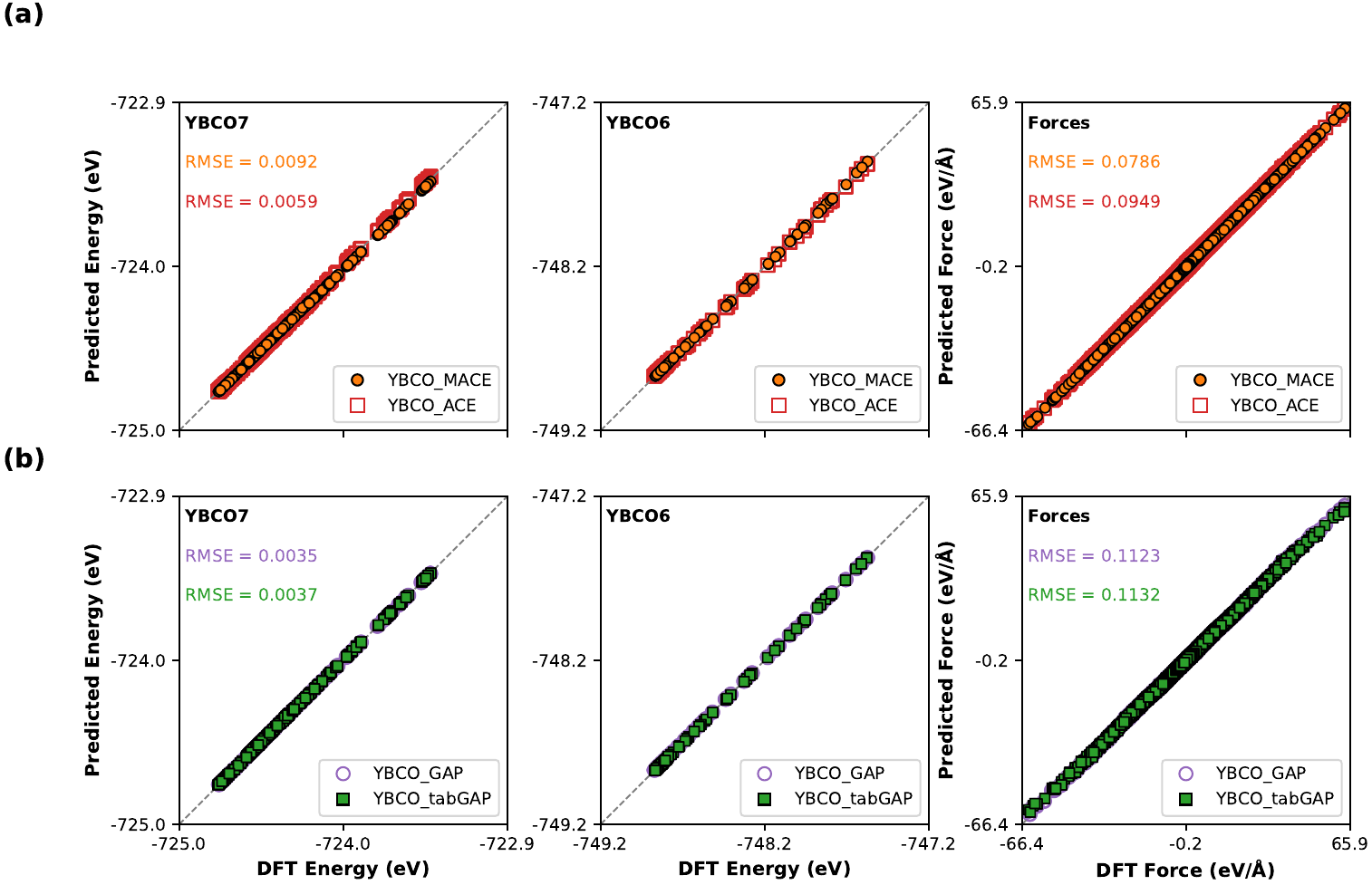}
    \caption{Parity plots for YBCO$_7$ (left) energy, YBCO$_6$ (center) energy, and forces (right) predicted by (a) YBCO\_MACE and YBCO\_ACE and by (b) YBCO\_GAP and YBCO\_tabGAP. RMSE values over the whole validation set for both energies and forces are included with the models' corresponding colors.}
    \label{fig:parplot}
\end{figure*}

As a first validation of our MLPs, we have constructed parity plots for energy and forces of our MLPs compared with an independent set of validation data. This data was not used for training. The validation set contains a wide range of volumes, thermal perturbations, and strains for YBCO$_7$ and YBCO$_6$. In total, $\sim$750 structures are used. The predicted energies and forces, together with the Root Mean Squared Error (RMSE), of our MLPs versus the reference DFT energies are shown in Fig.~\ref{fig:parplot}. The same color code for each MLP is employed throughout the whole manuscript.

Because the YBCO\_ACE model is trained on energies from which fixed per-element DFT reference values have been subtracted, its raw energies lie on a shifted internal scale and differ from the corresponding DFT energies by a composition-dependent constant. For information about rescaling, see Supplemental Material at~\cite{SuppMat}.

In all cases the energies and forces follow the DFT standard tightly. The energy errors are very similar between YBCO\_GAP and YBCO\_tabGAP, with YBCO\_ACE and YBCO\_MACE showing slightly worse errors. For YBCO\_MACE, this behavior arises from the two–stage training protocol employed in training. In this workflow, the first stage (\texttt{stage\_one}) establishes a more general and balanced model between energy and force accuracy, while the second stage (\texttt{stage\_two}) fine-tunes the energy alignment, sometimes at the expense of transferability and force precision. In the present work, we selected the \texttt{stage\_one} model as it yielded better results overall. The slightly increased energy RMSE of MACE therefore reflects a trade-off deriving from prioritizing force predictions and stability, that are crucial in radiation damage environments. Indeed, for forces YBCO\_MACE performs the best, followed by YBCO\_ACE, then the YBCO\_GAP and YBCO\_tabGAP.

\subsection{Bulk and Thermodynamic Properties}

\begin{figure*}[ht!]
    \centering
    \includegraphics[width=\linewidth]{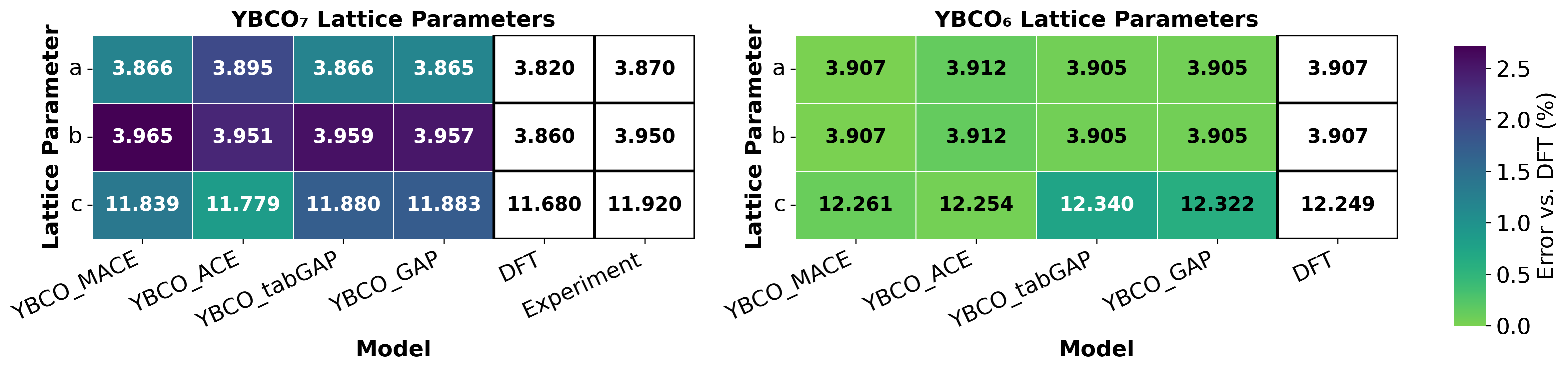}
    \caption{
        Comparison of lattice parameters (in~\AA) for YBCO$_7$ and YBCO$_6$ from
        YBCO\_MACE, YBCO\_ACE, YBCO\_tabGAP, YBCO\_GAP, DFT, and experiment~\cite{williams1988joint}.
        Experimental values are not included for YBCO$_6$ as they are not available.
        The lattice parameters are color-coded based on their error relative to the DFT ground truth.
    }
    \label{fig:ybcoparams}
\end{figure*}

Figure~\ref{fig:ybcoparams} compares the equilibrium lattice parameters predicted by YBCO\_MACE, YBCO\_ACE, YBCO\_tabGAP, YBCO\_GAP, and DFT for both YBCO$_7$ and YBCO$_6$. Also shown are the experimental values, which are reproduced well by the DFT. All MLPs reproduce both the orthorhombic distortion in YBCO$_7$ ($a \neq b$) as well as the tetragonal symmetry of YBCO$_6$. This demonstrates that all MLP models are capable of correctly capturing the symmetry-breaking due to oxygen stoichiometry. The results for YBCO$_6$ are very well captured by all MLPs, with the only notable difference being the YBCO\_GAP and YBCO\_tabGAP for the $c$ parameter. On the other hand, a bigger discrepancy with respect to DFT can be noted in the case of YBCO$_7$. All the MLPs show larger errors for the $b$ axis, and YBCO\_ACE shows a shift in the $a$ axis parameter. The $c$ axis errors are comparable to those for the other lattice parameters. We attribute the slightly lower accuracy of the YBCO$_7$ lattice parameters to the larger number of defected configurations in the training set, which likely introduced a small accuracy penalty. Overall, the difference in lattice parameters is, at maximum, less than $2.5\%$.

\begin{figure*}[ht!]
    \centering
    \vspace{-2mm}
    \includegraphics[width=\textwidth]{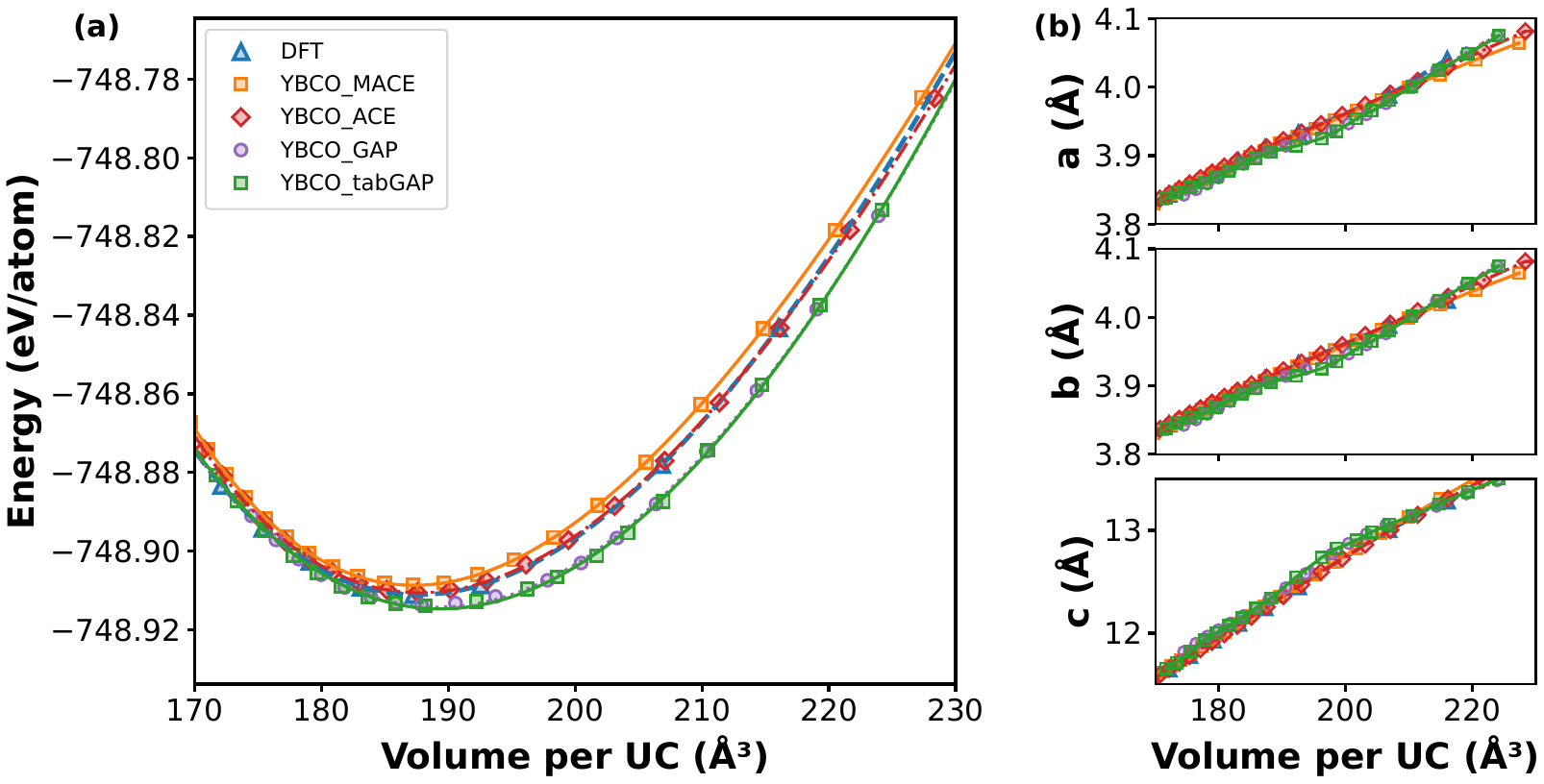}
    \vspace{-2mm}
    \includegraphics[width=\textwidth]{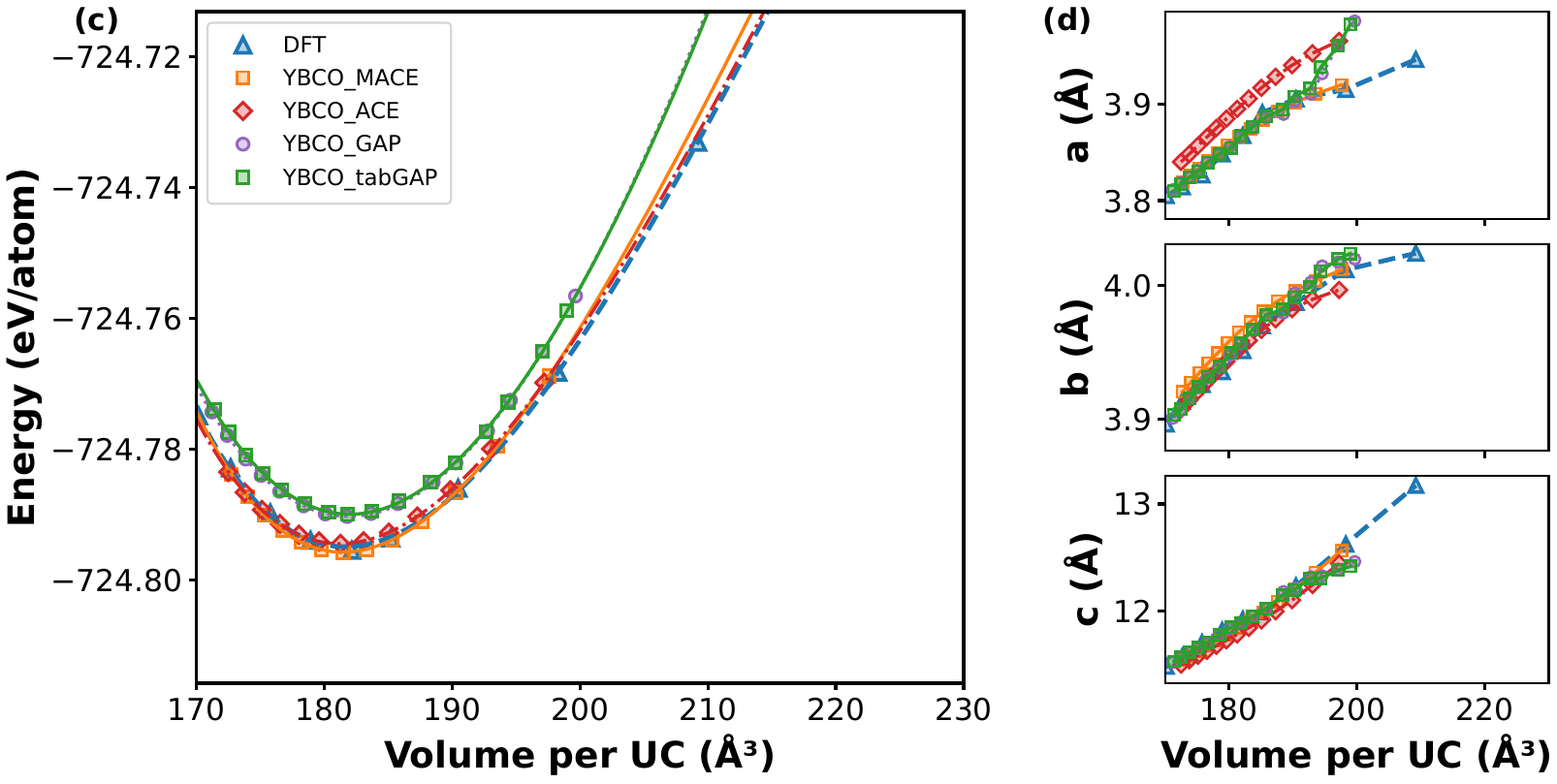}
    \caption{Comparison of the energy–volume curves and lattice parameters for YBCO$_6$ (top) and YBCO$_7$ (bottom) using DFT, YBCO\_MACE, YBCO\_ACE, YBCO\_GAP, and YBCO\_tabGAP models.}
    \label{fig:ybco_all}
\end{figure*}

The volume-energy curves for both YBCO$_6$ and YBCO$_7$ are presented in Fig.~\ref{fig:ybco_all} (a) and Fig.~\ref{fig:ybco_all} (c), respectively. These plots represent the change in energy as a function of lattice volume, obtained by minimizing the energy of the simulation cell under varying internal and external pressures (cell parameters were allowed to vary anisotropically). Accurate results here are crucial, as good agreement for the equation of state indicates that the system will respond correctly to lattice distortions. 

In all cases, the volume-energy curves obtained with the MLPs track the DFT results closely across a broad range of compressions and expansions. The curves were fitted using the third-order Birch–Murnaghan equation of state~\cite{Birch1947}, allowing a direct and consistent comparison between DFT and MLP energies and equilibrium volumes. YBCO\_MACE and YBCO\_ACE show a nearly perfect fit to the DFT data, whereas YBCO\_GAP and YBCO\_tabGAP exhibit some deviations. The predicted equilibrium volumes agree well with DFT, indicating that the MLPs not only reproduce interatomic forces accurately but also capture the underlying energetics governing structural stability.

In Fig.~\ref{fig:ybco_all} (b) and Fig.~\ref{fig:ybco_all} (d), lattice parameters values are shown as a function of unit-cell volume, for YBCO$_6$ and YBCO$_7$, respectively. It is noteworthy that at large volumes, YBCO\_GAP and YBCO\_tabGAP deviate from the DFT data for YBCO$_7$; however, this occurs for expansions exceeding approximately $+10\%$, and thus these discrepancies are not expected to significantly affect the performance of the model in the regime of interest. In contrast, YBCO\_ACE and YBCO\_MACE follow the DFT lattice parameter trends closely, with YBCO\_ACE showing a small systematic shift in the $a$ parameter. In general, YBCO\_MACE appears as the most accurate model as it shows very accurate fits even in the presence of lattice distortion.

\begin{figure*}[ht!]
    \centering
    \includegraphics[width=\textwidth]{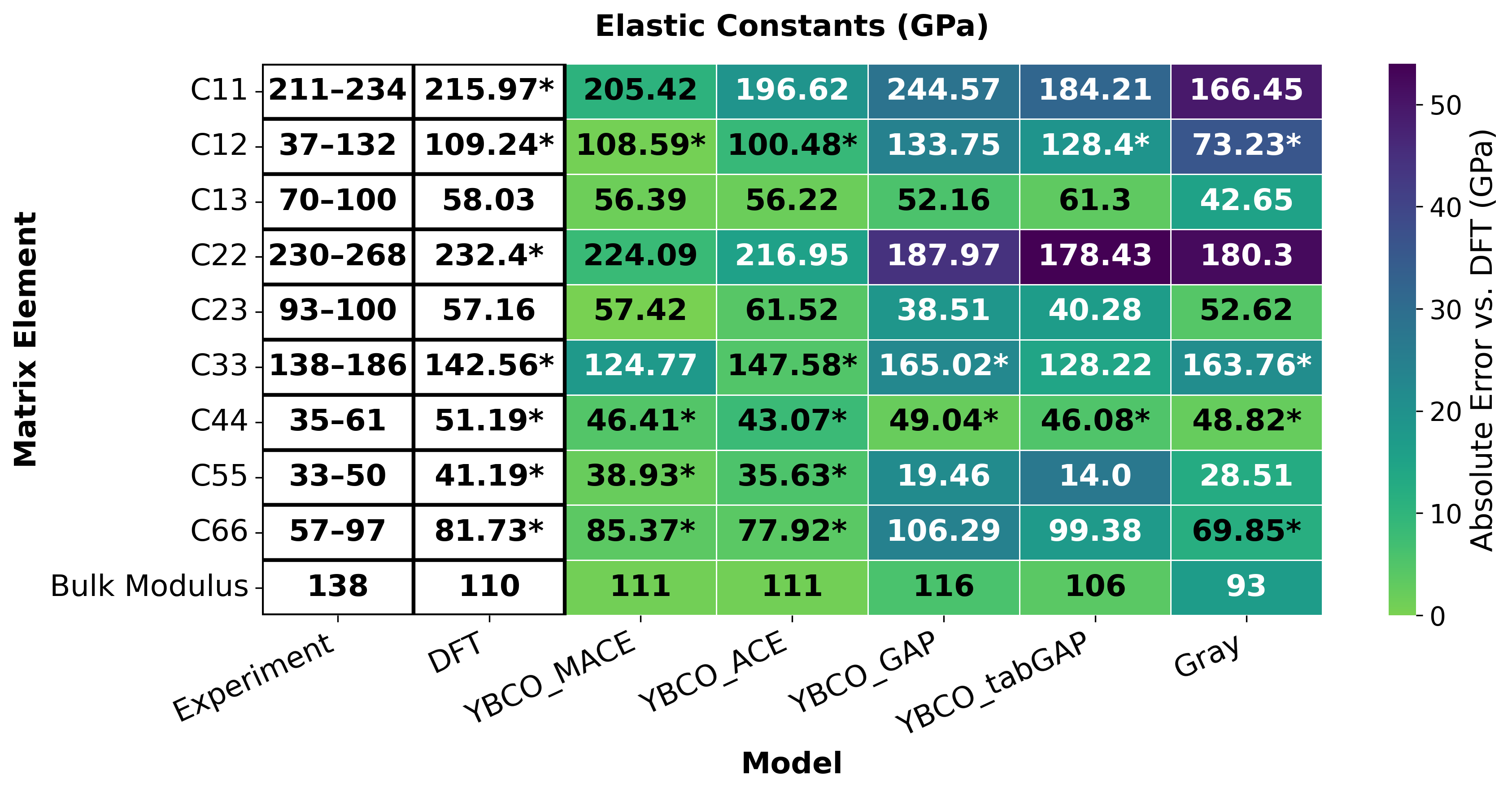}
    \caption{
        Comparison of elastic constants and bulk modulus values from
        DFT~\cite{murphy2020point}, the Gray potential, experiments~\cite{fernandes1991effect,lei1993elastic},
        YBCO\_MACE, YBCO\_ACE, YBCO\_GAP, and YBCO\_tabGAP.
        Values marked with an asterisk ($*$) fall within the experimental range.
    }
    \label{fig:elconst}
\end{figure*}

The calculated elastic constants and bulk moduli for all MLPs are shown in Fig.~\ref{fig:elconst}. The predicted bulk moduli of all MLPs are closer to both DFT and experimental values than the Gray model. We note that DFT itself may produce imperfect bulk moduli due to the treatment of electron exchange. In particular, the strongly correlated Cu 3$d$ electrons are poorly described by GGA functionals; in the absence of a Hubbard $U$ correction \cite{Liechtenstein1995, Dudarev1998} (or hybrid functional), electron localization is underestimated, weakening Cu–O bonds and thereby reducing the bulk modulus. Nevertheless, the agreement with DFT is good, and as this represents the reference ground truth, the MLPs cannot be expected to outperform it.

Consistent with the equation-of-state results, the bulk moduli of the YBCO\_ACE and YBCO\_MACE models are an excellent match to the DFT results. The YBCO\_GAP and YBCO\_tabGAP give a comparatively poor result for individual elastic constants, whereas the overall bulk moduli are good. Interestingly, the C$_{22}$ elastic constant proves to be the most challenging to reproduce.

\begin{figure*}[ht!]
    \centering
    \includegraphics[width=\textwidth]{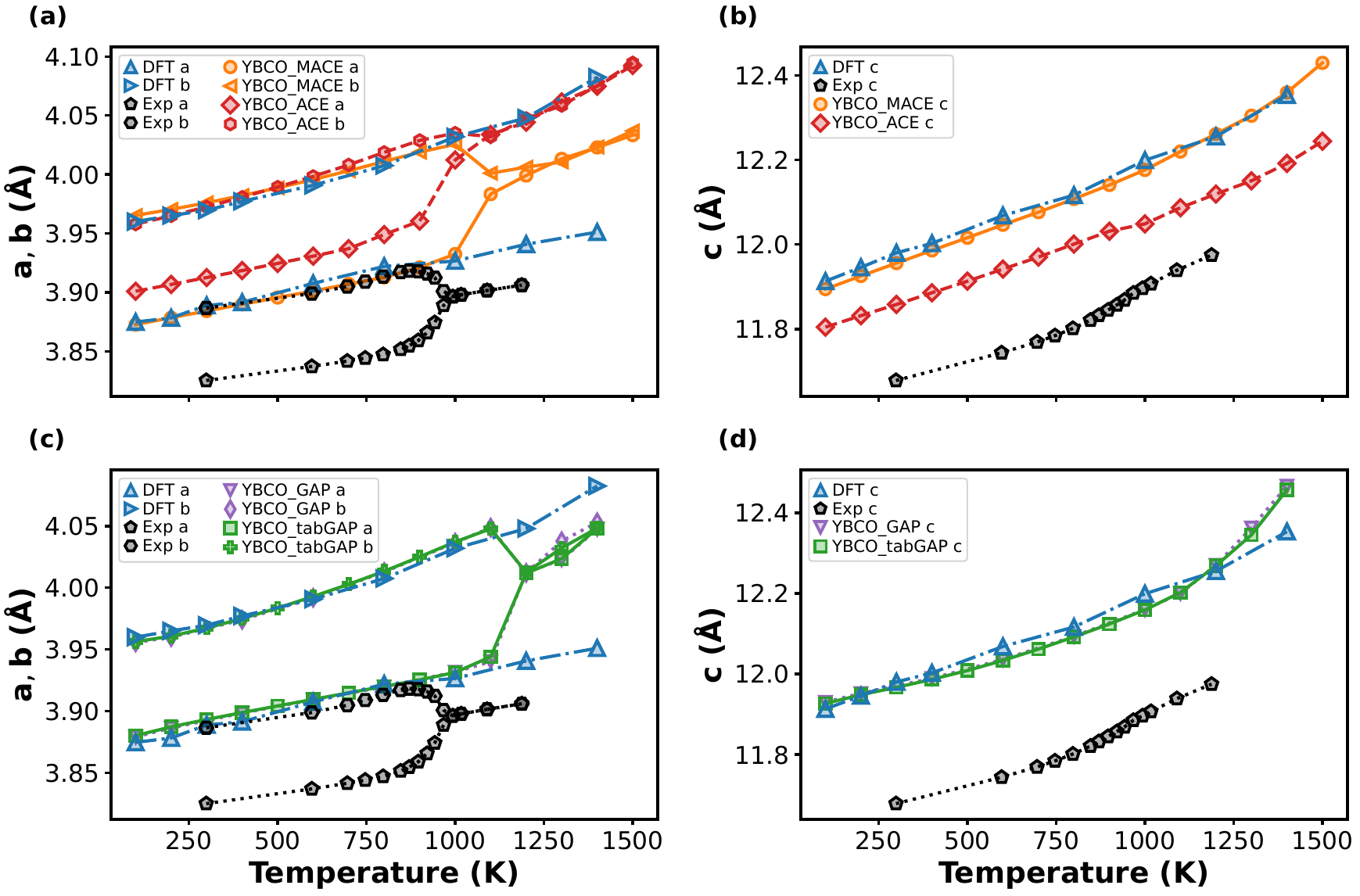}
    \caption{Thermal expansion of the lattice parameters in YBCO calculated with DFT, MACE, ACE, GAP, tabGAP, and experiment~\cite{Jorgensen1987}.}
    \label{fig:thermexp}
\end{figure*}

Finally, we examine the thermal expansion of YBCO predicted by our potentials, in comparison with DFT and experimental data from Ref.~\cite{Jorgensen1987} (see Fig.~\ref{fig:thermexp}). The agreement between the MLPs and DFT is excellent. As temperature increases, oxygen migrates to interstitial positions adjacent to the CuO chains, driving an orthorhombic-to-tetragonal phase transition. This transition is inherently kinetically driven and is therefore completely impractical to access via Born-Oppenheimer Molecular Dynamics simulations, which are limited to much smaller timescales. Conversely, our MLPs accurately reproduce this behavior, directly demonstrating their capability to model longer-timescale, temperature-activated phenomena at near-DFT accuracy. This thus represents a result that would otherwise be computationally unfeasible with first-principles methods.

Extending the simulation time appears to bring the phase transition closer to experimental behavior, but this lies beyond the scope of the present study and has already been shown in Ref.~\cite{Gambino2025}. The timescale required to observe the phase transition varies between models, primarily due to differences in the migration energies of chain oxygen. Notably, there is a systematic shift in the YBCO\_ACE $a$ and $c$ parameters relative to DFT. Nevertheless, the temperature dependence of thermal expansion predicted by all MLPs closely follows the trends observed in both DFT and experiment, demonstrating that the models capture the thermomechanical response of YBCO with high fidelity over extended timescales.

\subsection{Defects}

The introduction of defects during radiation damage cascades is directly relevant to the superconducting properties of YBCO tapes~\cite{fischer2018effect}. Consequently, it is important to monitor the performance of MLPs in predicting the energies of different defect types in YBCO. Due to the challenges associated with experimentally determining defect energies, we compare our results to DFT-calculated values obtained from CP2K.

\begin{figure*}[ht!]
    \centering
    \includegraphics[width=\linewidth]{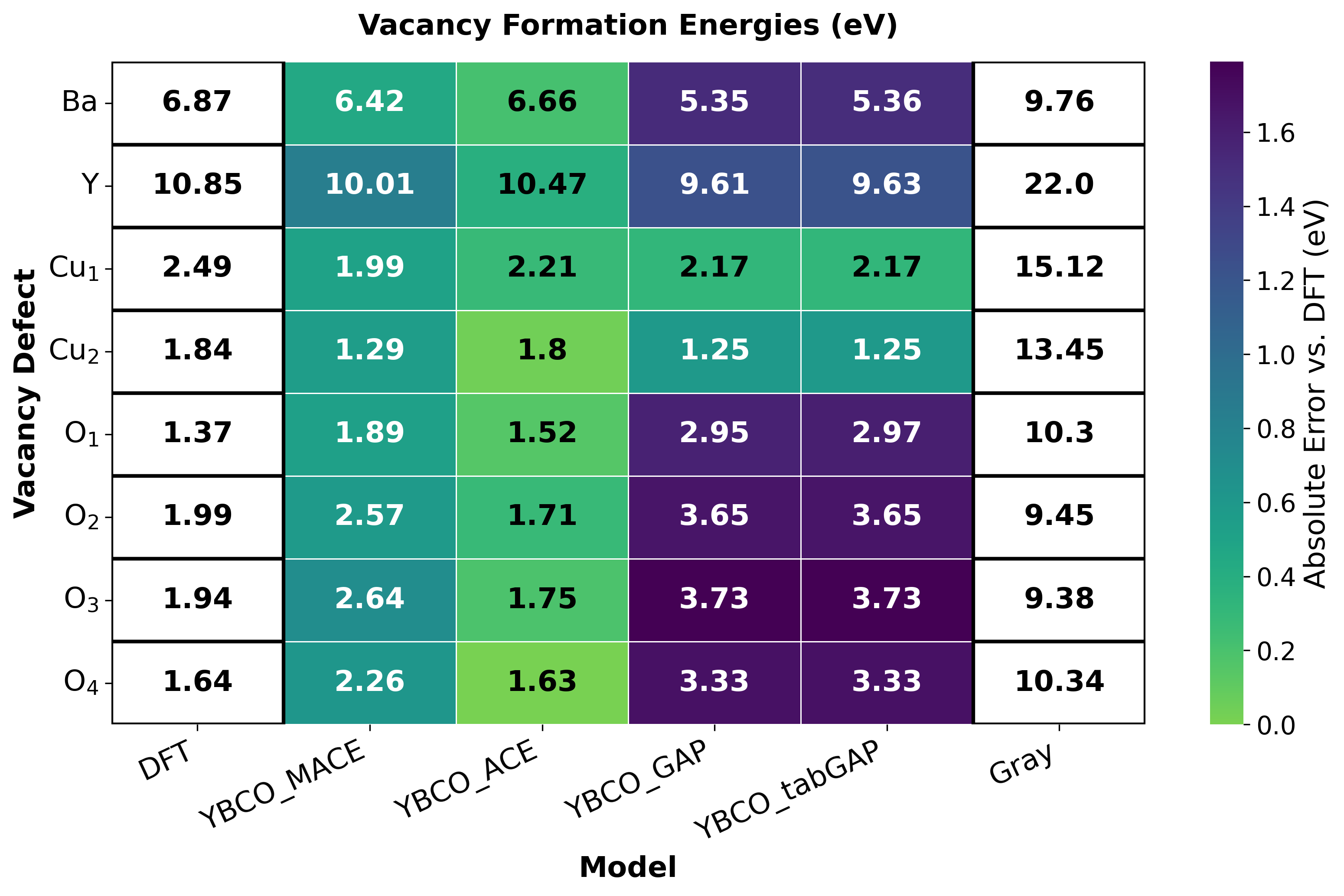}
    \caption{
        Vacancy defect formation energies predicted by
        YBCO\_MACE, YBCO\_ACE, YBCO\_GAP, YBCO\_tabGAP, DFT, and the Gray potential~\cite{gray2022molecular}.
        The discrepancy between machine-learned potential (MLP) predictions and the DFT standard
        is indicated by color.
    }
    \label{fig:vacancy}
\end{figure*}

A key advantage of MLPs over previous models for YBCO is their ability to be explicitly fitted to chemical potentials derived from DFT, which in principle allows direct comparison between DFT and the potentials. In practice, however, this is challenging due to the large number of degrees of freedom in YBCO. Indeed, all of our models struggled to accurately capture the chemical potential of oxygen in particular (see Supplemental Material at~\cite{SuppMat}), with the largest discrepancy observed for YBCO\_GAP/YBCO\_tabGAP. Similar issues have been reported previously, for example in the Ga$_2$O$_3$ tabGAP potential~\cite{zhao2023complex}, where deviations were attributed to the limited flexibility of low-dimensional GAP/tabGAP descriptors. This explanation is plausible here, since all MLPs were trained on the same database, yet the more flexible YBCO\_ACE and YBCO\_MACE potentials reproduce the chemical potential more accurately than YBCO\_GAP. The challenge is further compounded by the extreme diversity of local oxygen environments in complex oxides, which makes it difficult to capture all properties accurately. Additionally, the large number of distinct oxygen environments in the training set likely outweighs the inclusion of O$_2$ dimers, limiting the ability of the models to reproduce the O$_2$ reference.

The calculated vacancy energies for all potentials are shown in Fig.~\ref{fig:vacancy}. The vacancy defect energies predicted by the potentials exhibit varying levels of agreement with CP2K DFT results, with the largest deviations observed for oxygen vacancies, consistent with the discussion above. Our MLP results align with DFT predictions that the O1 site corresponds to the lowest-energy oxygen vacancy~\cite{liu2023first, Dickson2026}. By contrast, Murphy~\cite{murphy2020point} reports the O4 site as the most favorable, followed by O1. However, the difference is on the order of $0.1$ eV which could be attributed to DFT finite size error. Although it cannot be clearly determined which ordering of the O$_1$ and O$_4$ vacancies is correct, our MLPs capture the same trends as the training data. This represents a significant improvement over the Chaplot~\cite{Chaplot1989} and Gray et al.~\cite{gray2022molecular} potentials, which show the plane oxygens (O$_2$ and O$_3$) as the most favorable vacancy defects (in opposition to all prior DFT evidence). The Baetzold~\cite{baetzold1988atomistic} potential does correctly predict the ordering of the O vacancies, however, the copper vacancy order is instead reversed. Therefore, for vacancy defect energies, our MLPs represent an improvement over all previous models.

For oxygen vacancies, the YBCO\_ACE and YBCO\_MACE potentials outperform YBCO\_GAP and YBCO\_tabGAP. They also yield improved predictions for Y and Ba vacancies, which are systematically underpredicted by YBCO\_GAP and YBCO\_tabGAP. All potentials reproduce the ordering and energies of copper vacancies accurately, despite the different charge states of the Cu atoms. Overall, YBCO\_ACE provides the most consistent and accurate predictions across all vacancy defect types. For completeness, we also include the vacancy energies from the Gray potential, which is the only other potential made for radiation damage analysis in YBCO. These are not colored based on error from DFT as the Gray potential cannot be fit to chemical potentials, meaning the energy of defects is calculated simply as the energy difference between the defect cell and the perfect cell. However, the ordering of the vacancies can be directly compared.

\begin{figure*}[ht!]
    \centering
    \includegraphics[width=\linewidth]{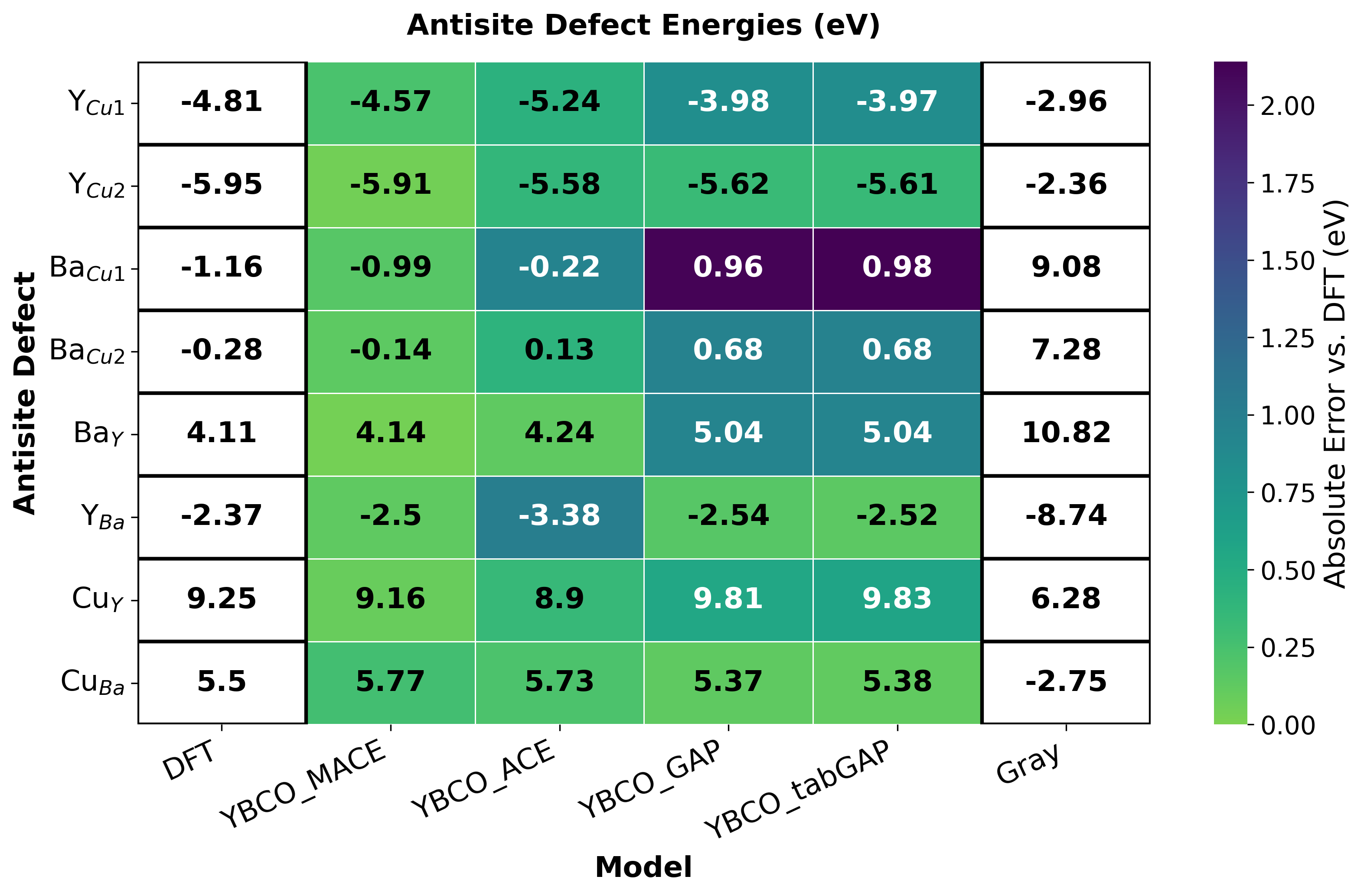}
    \caption{
        Antisite defect formation energies predicted by
        YBCO\_MACE, YBCO\_ACE, YBCO\_GAP,
        YBCO\_tabGAP, DFT, and the Gray potential.
        The deviation between machine-learned potential (MLP) predictions and the DFT standard
        is indicated by color.
    }
    \label{fig:antisites}
\end{figure*}

Next, we examine antisite defects in YBCO, as shown in Fig.~\ref{fig:antisites}. We use the standard Kr\"oger--Vink defect notation \(A_{B}\) \cite{Kroger1956}, in which an atom of species \(A\) occupies a crystallographic site normally associated with species \(B\). For example, \(\mathrm{Ba}_{\mathrm{Cu1}}\) denotes a barium atom substituting onto a Cu1 site in the YBCO lattice.

In general, the agreement between DFT values and the potential predictions is good. Again, YBCO\_ACE and YBCO\_MACE perform better than YBCO\_GAP and YBCO\_tabGAP, although for most antisite defects the predictions are similarly accurate. The Ba$_{\mathrm{Cu1}}$ defect is most poorly predicted by all but YBCO\_MACE. This is likely due to the level of distortion required to accommodate this defect, pushing the potentials into an extrapolative regime. In this scenario the YBCO\_MACE potential clearly performs better as the most expressive model. The Gray potential performs very well for the Ba$_{\mathrm{Y}}$ and Ba$_{\mathrm{Cu2}}$ defects, however, its performance its relatively for the Y$_{\mathrm{Cu1}}$ and Y$_{\mathrm{Cu2}}$ defects.

\begin{figure*}[ht!]
    \centering
    \includegraphics[width=\linewidth]{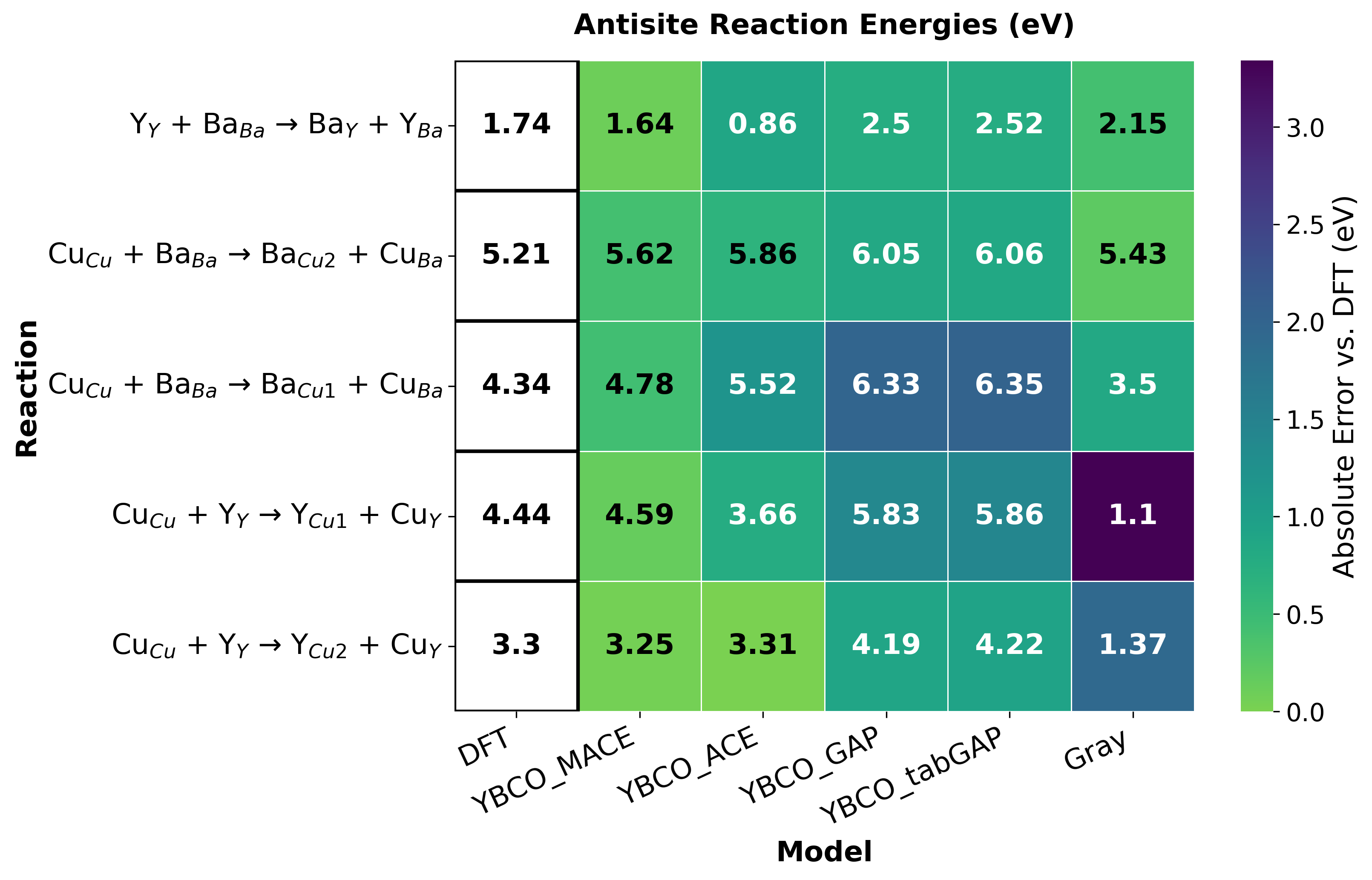}
    \caption{
        Reaction energies for antisite defects predicted by
        ACE, MACE, GAP, tabGAP, DFT, and the Gray potential.
        The deviation between machine-learned potential (MLP) predictions
        and the DFT standard is indicated by color.
    }
    \label{fig:reactions}
\end{figure*}

In Fig.~\ref{fig:reactions}, we also present reaction energies for various antisite exchange processes. One of the more favorable defect processes, according to Murphy~\cite{murphy2020point}, is the exchange of cations (Y and Ba), so it is important that this process is captured correctly. Due to the cancelling of chemical potentials here, the Gray potential can be directly compared to the MLPs. The exchange process is captured well by all potentials, including Gray's. The accuracy of the latter does however fall short for the other antisite processes involving copper, where the MLPs show superior results. Again, the predictions by YBCO\_ACE and YBCO\_MACE are better than those of YBCO\_GAP and YBCO\_tabGAP. The exchange of Ba and Y is reasonably reproduced by YBCO\_GAP and YBCO\_tabGAP, although the predicted energies are slightly higher than DFT.

\begin{figure*}[ht!]
    \centering
    \includegraphics[width=\linewidth]{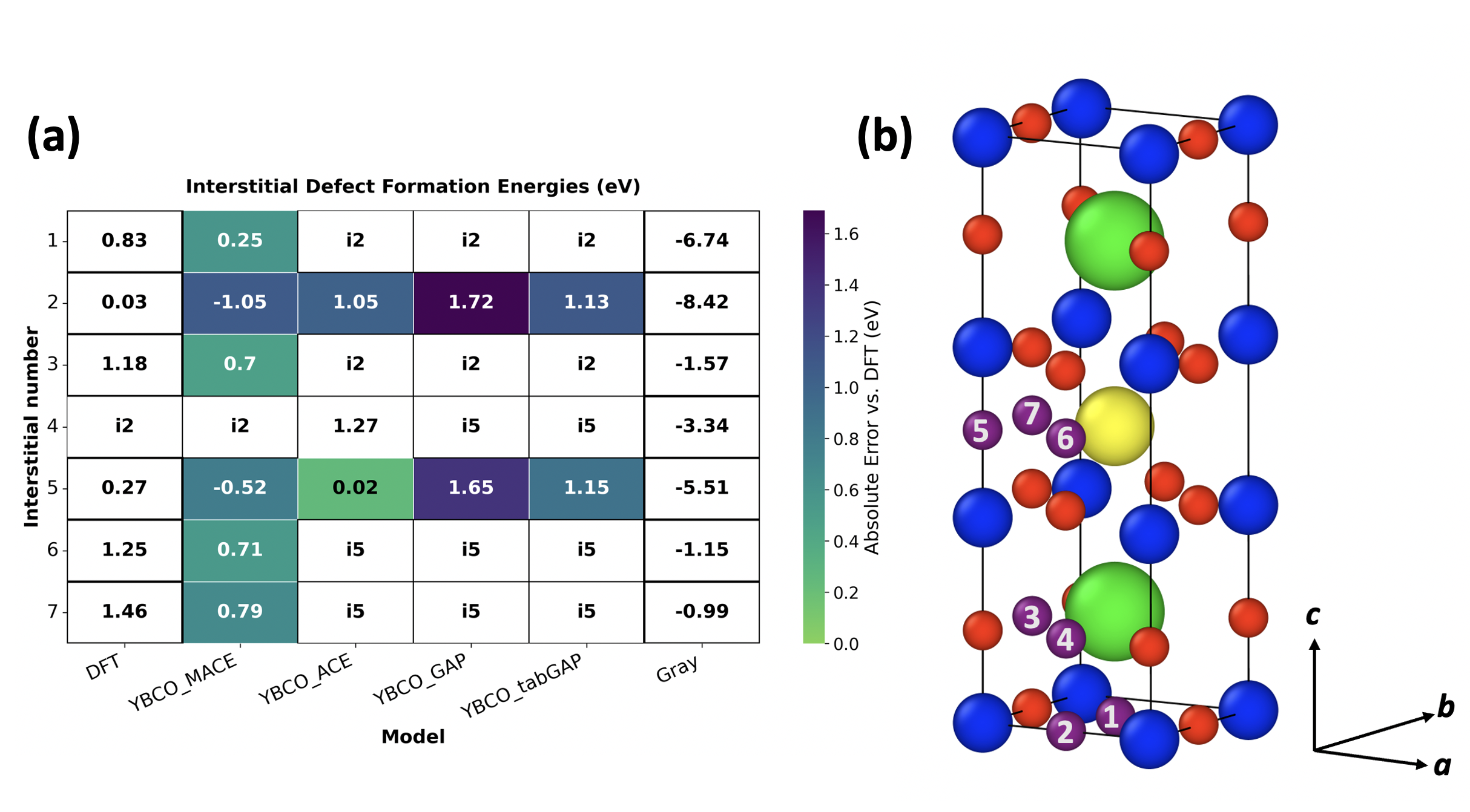}
    \caption{(a) Oxygen interstitial defect formation energies predicted by YBCO\_ACE, YBCO\_MACE, YBCO\_GAP, YBCO\_tabGAP, DFT, and the Gray potential. The error between MLP predictions and the DFT standard is shown by color. When an interstitial site relaxes into another configuration, this is denoted by ``i'' followed by the index of the site into which it collapses. (b) Interstitial sites in YBCO.}
    \label{fig:interstitials}
\end{figure*}

To conclude our discussion on single-defect formation energies, we examined the oxygen interstitial configurations in YBCO, identified as the most mobile species, following the sites reported by Murphy~\cite{murphy2020point}. These are summarized in Fig.~\ref{fig:interstitials}. When an interstitial site relaxes into another configuration, this is denoted by ``i'' followed by the index of the site into which it collapses. Our DFT calculations predict that the interstitial site~2 is the most stable configuration, followed by site~5. In contrast to the results of Murphy~\cite{murphy2020point}, our DFT calculations yield peroxide-like split interstitial configurations for sites~6 and~7, whereas Ref.~\cite{murphy2020point} reports these sites as collapsing back to site~5.

The MLP predictions show notably larger variations. Since all configurations outside of sites~2 and~5 in our DFT results form split interstitials characterized by short O--O bond distances, accurately reproducing these defects requires a well-described oxygen chemical potential. Due to the known inaccuracies in the oxygen chemical potential for YBCO\_GAP and YBCO\_tabGAP, the resulting defect landscape is simplified, with all interstitials relaxing to either site~2 or site~5. Interestingly, while YBCO\_tabGAP reproduces the correct energetic ordering of sites~2 and~5, YBCO\_GAP does not, though slight differences in relaxed geometries could account for this discrepancy. 

YBCO\_ACE produces a simplified defect landscape: all sites except interstitial~4 collapse to other configurations in the same manner as YBCO\_GAP and YBCO\_tabGAP. The interstitial~4 site identified by YBCO\_ACE forms a split interstitial configuration not observed in our DFT results. YBCO\_ACE reverses the energetic ordering of sites~2 and~5, with a substantial energy difference between them. In contrast, YBCO\_MACE correctly reproduces the DFT standard: both the ordering of sites~2 and~5 and the relaxation of site~4 into site~2. YBCO\_MACE also identifies distinct defect configurations for all other interstitial sites, in agreement with DFT. 

It is worth noting that, since the DFT reference energies for interstitials are very low and the oxygen chemical potential is inaccurate compared to the DFT standard, the combination of these two factors can lead to negative formation energies. These negative values should not however be interpreted as implying spontaneous defect formation, but rather as the relative stability of the possible interstitial configurations.

It can thus be concluded that, in general, MACE is the most accurate MLP formalism for calculating defects' formation energies in YBCO$_7$. It is worth mentioning that there is little experimental evidence for the formation of peroxide-like defects in metal oxides. There is some evidence in zirconate perovskites~\cite{middleburgh2014peroxide}, however, considering how common these defects are in GGA DFT simulations~\cite{youssef2012intrinsic, erhart2005first, murphy2013anisotropic}, one would expect more evidence of their existence. It is therefore unclear whether capturing these interstitial defect properties is necessary for accurate simulations.

\begin{figure*}[ht!]
    \centering
    \includegraphics[width=\textwidth]{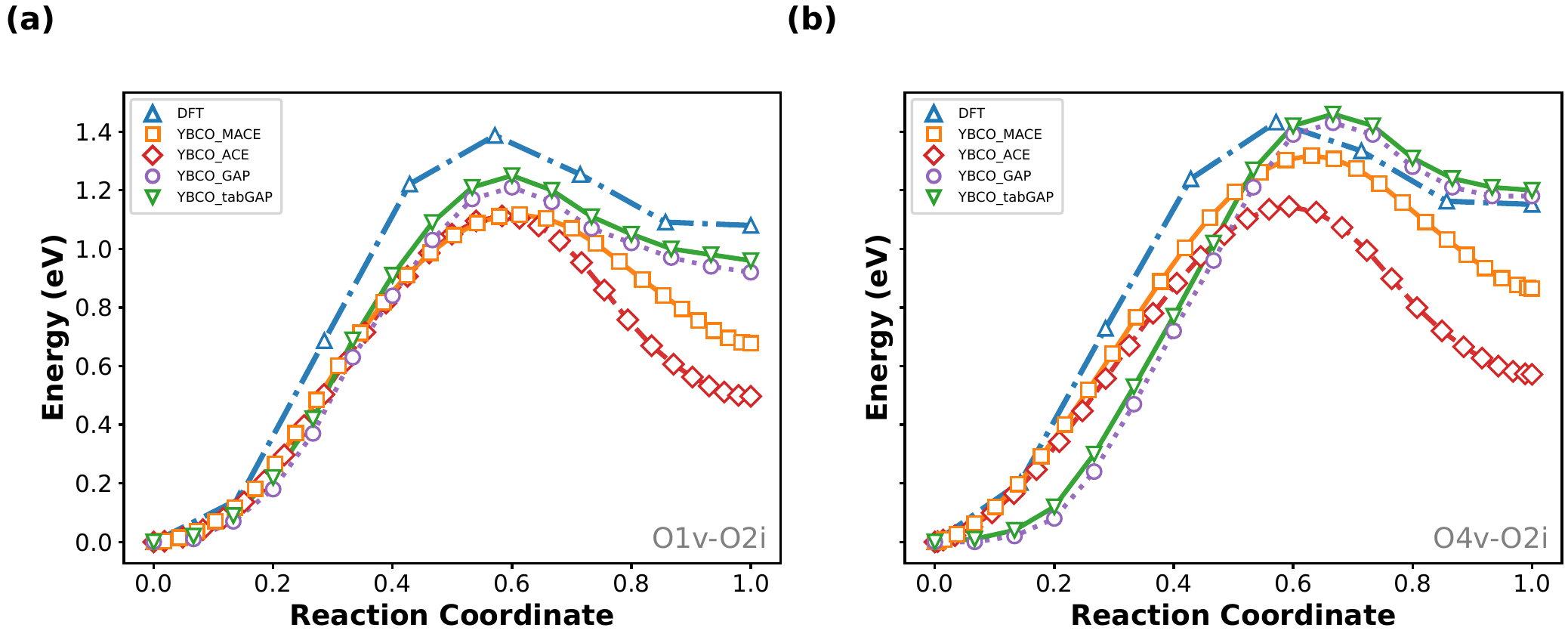}
    \caption{NEB energy profiles for (a) migration of oxygen~1 to interstitial position~2 and (b) oxygen~4 to interstitial~2. Results are shown for DFT, YBCO\_MACE, YBCO\_ACE, YBCO\_GAP, and YBCO\_tabGAP models.}
    \label{fig:neb}
\end{figure*}

As a final validation of defect properties, we calculated activation barriers of oxygen migration using climbing-image NEB, as shown in FIg. \ref{fig:neb}. This approach allows us to probe the energy pathways for two defect migration processes: migration of a chain O1 atom to interstitial position~2, and migration of an apical O4 atom to the same interstitial site. These atoms are expected to be the most mobile, consistent with their defect energies, and likely play a key role in the orthorhombic-to-tetragonal phase transition. This has been elucidated in a computational study by Gambino \textit{et al.}~\cite{Gambino2025}. We compare the NEB pathways calculated with our potentials to those obtained from DFT. It is important to mention that NEB DFT calculations in CP2K were performed on a $4\times4\times2$ YBCO$_7$ supercell, due to convergence issues.

The predicted reaction energies for both pathways are similar, approximately $1.4$~eV. For both reactions, YBCO\_GAP and YBCO\_tabGAP show excellent agreement with the DFT results. While the transition state for the O4 atom migration to the interstitial 2 position is located at a slightly different reaction coordinate, this discrepancy is likely due to the limited number of images used in the DFT calculations. YBCO\_ACE and YBCO\_MACE show somewhat larger deviations from the DFT reference, with YBCO\_ACE in particular exhibiting a notable discrepancy. Overall, the close agreement between the machine-learned potentials and DFT in these NEB calculations provides strong validation for the modeling of oxygen migration in YBCO.

\subsection{Handling oxygen stoichiometry with MLPs}

\begin{figure*}[ht!]
    \centering
    \includegraphics[width=\textwidth]{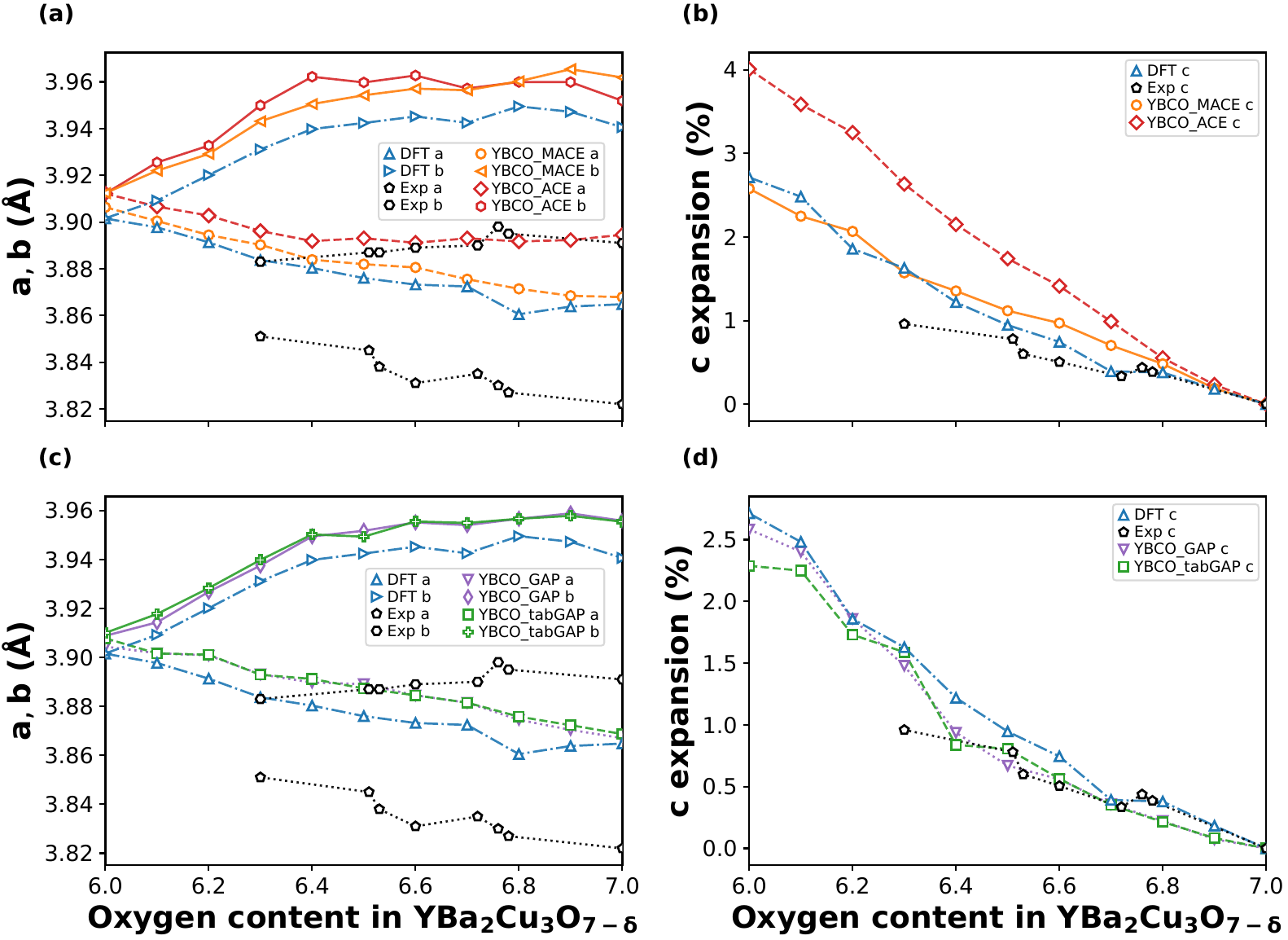}
    \caption{
        Comparison of lattice parameter expansions as a function of oxygen content in YBa$_2$Cu$_3$O$_{7-\delta}$.
        Results are shown for (a) $a,b$ lattice parameters as a function of oxygen content for DFT, YBCO\_MACE, YBCO\_ACE, and experimental data;
        (b) $c$ lattice parameter percentage expansion as a function of oxygen content for DFT, YBCO\_MACE, YBCO\_ACE, and experimental data;
        (c) $a,b$ lattice parameters as a function of oxygen content for DFT, YBCO\_GAP, YBCO\_tabGAP, and experimental data;
        (d) $c$ lattice parameter percentage expansion as a function of oxygen content for DFT, YBCO\_GAP, YBCO\_tabGAP, and experimental data. All experimental data is from \cite{Jorgensen1987}.
    }
    \label{fig:lps_vs_O_all}
\end{figure*}

One of the principal strengths of MLPs lies in their ability to learn from broad and diverse datasets, allowing them to accurately represent both stoichiometric variations in complex materials and the highly disordered atomic configurations that can be found in collision cascades. Figure~\ref{fig:lps_vs_O_all} compares the lattice parameters predicted by our MLPs with experimental and DFT results across a range of oxygen contents (from O$_7$ to O$_6$). The DFT-predicted expansion of the $c$-axis shows excellent agreement with experiment, and all MLPs closely reproduce this behavior, with the exception of the YBCO\_ACE model, which deviates significantly at lower oxygen contents. The YBCO\_ACE potential also exhibits larger discrepancies from DFT for the $a$ and $b$ lattice parameters compared to the other models.

Oxygen removal causes the $a$ and $b$ lattice parameters to converge, leading to tetragonal symmetry at an oxygen concentration of~6, a trend that is consistent with experimental observations~\cite{Jorgensen1987} and accurately reproduced by all MLPs. This agreement provides strong evidence of the robustness and transferability of the potentials to oxygen-deficient compositions. Furthermore, obtaining accurate prediction of lattice parameter changes as a function of oxygen content demonstrates the ability of these MLPs to capture the coupling between structure and stoichiometry. 

It is worth noting that the difference between experimental and MLP lattice parameters is due to the DFT training data. DFT GGA functionals (such as the PBE~\cite{perdew1996generalized} functional used in this work) tend to under-bind~\cite{yuk2024putting}, therefore leading to an overestimation of bond lengths. This can in general be counteracted by using higher levels of theory, however, the associated computational cost of these functionals can make it challenging to create a diverse dataset for an MLP.

\subsection{Performance Benchmarking}\label{bench}

We benchmarked the computational efficiency of the MLPs developed in this work to provide guidance on selecting the most suitable potential for a given application. All benchmarks were performed with LAMMPS (version 12Jun-2025). We do not recommend using the YBCO\_GAP potential, as the YBCO\_tabGAP potential is significantly faster while maintaining comparable accuracy; consequently, YBCO\_GAP was not included in these benchmarks.

Given the comparable computational performance of the YBCO\_ACE and YBCO\_tabGAP potentials, it was essential to evaluate them under identical conditions. CPU benchmarks were performed in the NVE ensemble on a 936-atom YBCO$_7$ supercell using a dual-socket AMD EPYC~9334 system. Each compute node consists of 2~$\times$~AMD EPYC~9334 32-core processors (2.7\,GHz), providing a total of 64~physical cores (128~hardware threads) arranged across 2~NUMA domains. Simulations were executed using 48~CPU cores for consistency with previous work.

GPU benchmarks for YBCO\_ACE and YBCO\_tabGAP employed one compute node consisting of four NVIDIA A100~SXM-80GB GPUs and 2~$\times$~AMD EPYC~7763 64-core processors (1.8\,GHz). We used LAMMPS (with the Kokkos~\cite{kokkos1, kokkos2} package) compiled with CUDA and OpenMPI support. Jobs were launched using Slurm with 4~MPI ranks per node, corresponding to one MPI rank bound to each GPU, and 32~CPU cores per rank. Simulation cells contained 3,833,856~atoms, and GPU benchmarks were performed in the NVE ensemble, matching the CPU setup.

Given that the scaling and general performance of YBCO\_MACE is significantly slower than YBCO\_ACE and YBCO\_tabGAP (see the scaling results in \cite{Kovacs2025, batatia2025foundation}), we have performed the YBCO\_MACE GPU benchmark on a smaller system of $7,488$ atoms. Simulations were once again performed in NVE MD, using the same hardware and Slurm setup as described above. The YBCO\_MACE simulations utilised the \verb|symmetrix| \cite{Kovacs2025, batatia2025foundation} evaluator and Kokkos \cite{kokkos1, kokkos2}.  Batatia \textit{et al.} \cite{batatia2025foundation} recommend the use of 1000-4000 atoms per GPU, therefore, our benchmark is representative of the (current) ideal performance of YBCO\_MACE.

The results indicate that YBCO\_MACE is significantly slower on CPU and GPU than YBCO\_ACE and YBCO\_tabGAP. Furthermore, the scaling of YBCO\_MACE with system size is less efficient. Given its superior performance on virtually all properties measured in this paper (with the exception of computational speed), YBCO\_MACE should be considered for use in smaller simulations where high accuracy is pertinent. Larger simulations (for instance, radiation damage focused work) should generally employ YBCO\_tabGAP and YBCO\_ACE, as their efficiency clearly outperforms YBCO\_MACE (YBCO\_tabGAP achieves almost the same number of steps for just under 4 million atoms than YBCO\_MACE does for just over $\sim7000$). YBCO\_tabGAP is roughly five times faster than YBCO\_ACE on GPU, and is over 50 \% faster on CPU.

\begin{table}[t]
\centering
\caption{Performance comparison of YBCO potentials. Benchmarks were performed in NVE MD with a 1 fs timestep. The ns/day metric indicates how much simulation time is passed in a day. CPU benchmarks were performed on 48 cores with 936 atoms for all models. GPU benchmarks used four NVIDIA A100 GPUs with $3\,833\,856$ atoms for YBCO\_ACE and YBCO\_tabGAP, and $7\,488$ atoms for YBCO\_MACE.
The asterisk indicates that the YBCO\_MACE GPU benchmark used a smaller cell.}
\label{tab:performance}

\begin{tabular}{lccc}
\hline \hline
\textbf{Hardware} & \textbf{YBCO\_MACE} & \textbf{YBCO\_ACE} & \textbf{YBCO\_tabGAP} \\
\hline
CPU (ns/day) & 0.90 & 33.84 & 53.24 \\
GPU (ns/day) & 0.56$^{*}$ & 0.09 & 0.46 \\
\hline \hline
\end{tabular}

\end{table}

\section{Conclusion}

In this work, four MLPs were fitted to an extensive DFT database for the YBCO system. For the first time, we produce reliable MLPs for analysis of bulk, defect, and thermodynamic properties in YBCO at a range of oxygen stoichiometries, which are relevant for radiation damage applications. Among these, the YBCO\_MACE model provides the most accurate results overall, albeit at a significantly higher computational cost. The YBCO\_ACE potential demonstrates improved performance relative to YBCO\_tabGAP and YBCO\_GAP for defect formation energies, stress tensors, and the equation of state. In contrast, the YBCO\_tabGAP and YBCO\_GAP yield more reliable results for representative NEB simulations and for predicting lattice parameters and stoichiometric variations in YBCO. The YBCO\_GAP and YBCO\_tabGAP potentials yield quantitatively similar predictions, and since YBCO\_tabGAP offers considerably better computational efficiency, we focus on YBCO\_tabGAP in the remainder of this discussion. The YBCO\_tabGAP performs well across most tested properties; however, its accuracy for oxygen-related defect energies is limited, primarily due to inaccuracies in the predicted oxygen chemical potential. However, we do not anticipate that this will have an effect on radiation damage results, as the simulations are performed far from equilibrium.\\

Overall, we recommend that smaller-scale studies focused on thermodynamic properties employ the YBCO\_MACE potential, given its higher accuracy. For larger-scale simulations, either YBCO\_ACE or YBCO\_tabGAP provide an appropriate balance between accuracy and efficiency, both being substantially faster than YBCO\_MACE. Since these models were developed with future radiation-damage simulations in mind, computational performance remains a key consideration. At present, YBCO\_tabGAP exhibits slightly better performance on CPU than YBCO\_ACE, and is $\sim$5 times more efficient on GPU architectures. Future work will therefore assess the comparative performance of these potentials for large-scale radiation-damage simulations.

\section*{Acknowledgements} \label{sec:acknowledgements}

DFT calculations have been performed by NDE and AD on, respectively, the JFRS-1 CRAY XC50 and the ARCHER2 CRAY XC30 HPCs. NDE trained and validated the YBCO\_MACE and YBCO\_ACE models, while AD trained and validated the YBCO\_GAP and YBCO\_tabGAP models. DG performed the VASP phonons simulations and generated both the initial structures for the lattice parameters vs. oxygen content simulations, as well as the amorphous ones.

Training and validations of YBCO\_MACE and YBCO\_ACE were enabled by Eni’s HPC4 and HPC6, while for YBCO\_GAP and YBCO\_tabGAP the CSD3 and the Lancaster University HEC were utilised. NDE would like to acknowledge Prof. Gábor Csányi for fruitful scientific discussions regarding MACE and Dr. W. Chuck Witt for precious support with the \verb|symmetrix| package. AD would like to thank Jesper Byggm\"astar for useful discussions.

NDE acknowledges that this publication is part of the project PNRR-NGEU which has received funding from the MUR – DM 117/2023. NDE, DT, FeL and FL acknowledge support from Eni S.p.A.. This project was also funded by Lancaster University and UKAEA via agreement 2077239. DNM and MRG acknowledge funding from the EPSRC Energy Programme [grant number EP/W006839/1]. DG acknowledges financial support from the Swedish Research Council (VR) through Grant
No. 2023-00208. This work was carried out (partially) using supercomputer resources provided under the EU-JA Broader Approach collaboration in the Computational Simulation Centre of International Fusion Energy Research Centre (IFERC-CSC).

\section*{Data Availability} \label{sec:data}

The data that support the findings of this article are openly available \cite{DiEugenio2025MACE, Dickson2025GAP}.

\bibliography{lib2}

\clearpage
\appendix
\onecolumngrid

\section*{Supplemental Material}

\section{ACE Rescaling}

As the YBCO\_ACE model is trained on energies from which fixed per-element DFT reference values have been removed, the printed raw energies differ from the DFT energies by a composition-dependent constant. We therefore apply a two-step correction. First, we remove the element-dependent bias in the YBCO\_ACE predictions by adding a composition-weighted linear correction. For a structure $i$ this gives
\begin{equation}
E_{\mathrm{ACE,bias\;corr}}^{(i)} =
E_{\mathrm{ACE}}^{(i)} +
\sum_{\alpha} n_{\alpha}^{(i)}\,\delta E_{\alpha},
\end{equation}
where $n_{\alpha}^{(i)}$ is the number of atoms of element $\alpha$ and
$\delta E_{\alpha}$ are per-element offsets obtained from a linear least-squares
fit to the DFT total energies:
\begin{equation}
\{\delta E_{\alpha}\}
=
\arg\min_{\{\delta E_{\alpha}\}}
\sum_{i}
\left[
E_{\mathrm{DFT}}^{(i)} -
E_{\mathrm{ACE}}^{(i)} -
\sum_{\alpha} n_{\alpha}^{(i)}\,\delta E_{\alpha}
\right]^{2}.
\end{equation}

Second, because the DFT energies used for training were shifted by subtracting the sum of their per-element reference energies, we restore the absolute DFT energy scale by adding this reference contribution back:
\begin{equation}
E_{\mathrm{ACE,DFTscale}}^{(i)} =
E_{\mathrm{ACE,bias\;corr}}^{(i)} +
\sum_{\alpha} n_{\alpha}^{(i)}\,E_{\mathrm{ref}}(\alpha).
\end{equation}
Both corrections depend only on composition, so forces and relative energies are unchanged.

As a worked example, consider a YBCO$_7$ supercell containing 936 atoms. The raw YBCO\_ACE energy is $E_{\mathrm{ACE}}=-5123.921400$~eV. The fitted per-element corrections give a composition-dependent shift of $-723.9970$~eV, yielding
\[
E_{\mathrm{ACE,bias\;corr}} = -5847.918400~\text{eV}.
\]
The per-element DFT reference energies for this composition sum to
$-672517.7694416$~eV, giving
\[
E_{\mathrm{ACE,DFTscale}}
=
-5847.918400
+
(-672517.7694416)
=
-678365.6883420~\text{eV}.
\]
Dividing by 936 atoms gives the final per-atom YBCO\_ACE value
\[
-724.749666497869~\text{eV/atom},
\]
which is then compared to the DFT result,
\[
-724.749115506002~\text{eV/atom}.
\]
This demonstrates how to bring back YBCO\_ACE energies onto the absolute DFT scale for individual configurations.

\section{Parameters of the ACE potential}
The ACE potential was constructed using a radial cutoff of 6.5~\AA\ and 700 basis functions per element for Ba, Cu, Y, and O. The model includes explicit high-body polynomial correlations up to sixth order for unary and binary terms and up to fifth order for ternary and quaternary terms, ensuring improved representation of local atomic environments. Simplified Bessel radial functions were employed along with a Finnis--Sinclair-type shifted and scaled embedding function, while an automatic short-range core repulsion was incorporated based on the Ziegler--Biersack--Littmark (ZBL) potential \cite{Ziegler1985}. The fitting was performed using a BFGS optimizer with a batch size of 100 and a maximum of 2000~iterations. The weighted loss function assigned relative weights of 10 for forces and 1 for energies, with $\kappa=0.3$. Overall, this ACE potential corresponds to an explicit linear expansion including up to seven-body terms within the high-body-order ACE formalism.

\section{Parameters of the MACE potential}

The MACE potential employs message passing with 128 scalar channels and $\mathrm{max\_L}=0$ (i.e., $128\times0e$), consisting of two interaction layers, each with correlation order three (corresponding to an effective four-body interaction range) and spherical harmonics up to $l=3$. The model utilizes eight radial and five angular basis functions, with a cutoff radius of 6.5~\AA, giving each atom a total receptive field of 13.0~\AA. The \texttt{Agnesi} distance transform is used for the radial basis functions, allowing a smooth transition based on atomic covalent radii, while a \texttt{pair\_repulsion} term introduces a short-range ZBL-like repulsive correction. Training was performed using a batch size of six and a weighted loss function, assigning weights of 10 for forces and 1 for energies. The optimizer was \texttt{AMSGrad} with gradient clipping of 5.0 and a weight decay of $5\times10^{-7}$, combined with exponential moving average (EMA, decay~0.99) and stochastic weight averaging (SWA) starting at epoch~425. The model was trained for up to 600~epochs on four distributed processes (MACE~v0.3.14). This setup corresponds to an equivariant interatomic potential capturing effective many-body correlations up to approximately four-body order.

\section{Parameters of the GAP potential}

The GAP potential employs three different descriptors: two-body, three-body, and EAM. Converged cutoff distances of 6.5 {\AA} are chosen for the two-body and EAM descriptors, and a cutoff of 4.5 {\AA} is chosen for the three-body. These choices provided a good balance of speed and accuracy. The number of sparse points was selected by determining the absolute maximum number we could use with our available hardware. This has a detrimental effect to the speed of the GAP. However, after tabulation, the number of sparse points has no effect on the tabGAP. Therefore we retain the higher resolution of the large number of sparse points, with no reduction in perfromance for the tabGAP. \\

The remaining hyperparameters were optimised using a Bayesian Optimisation approach (see section \ref{BOforGAP}). The final hyperparameters are shown in Supplementary table \ref{GAPparamstab}. In order to balance the contributions of forces and energies from different types of structures, we employed regularisation weights to different types of training data. The weights for the energies are as follows: 0.002 eV for solid YBCO$_{7}$/YBCO$_{6}$ (including defects), 0.01 eV for other crystal types, and 0.005 eV for liquid / amorphous structures. For the forces and virials, these weights were increased by a factor 10 in all cases. A screened coulomb repulsive potential was fitted to all-electron DFT data from \cite{nordlund2025repulsive}, and attached to the GAP. \\

\begin{table}[h!]
\centering
\caption{GAP parameters used in this work.}
\label{tab:gap_parameters}
\begin{tabular}{ll}
\hline\hline
\textbf{Descriptor} & Two-body \\ 
\hline
Cutoff (\AA)         & 6.5 \\
$\theta$             & 1 \\
$\delta$             & 2.97 \\
Sparse points        & 25 \\
Covariance type      & ARD\_SE \\
Sparse method        & Uniform \\
\hline
\textbf{Descriptor} & Three-body \\
\hline
Cutoff (\AA)         & 4.5 \\
$\theta$             & 2.37 \\
$\delta$             & 0.24 \\
Sparse points        & 550 \\
Covariance type      & ARD\_SE \\
\hline
\textbf{Descriptor} & EAM \\
\hline
Cutoff (\AA)         & 6.5 \\
$\theta$             & 1 \\
$\delta$             & 0.24 \\
Sparse points        & 20 \\
Pair function        & FSgen \\
Covariance type      & ARD\_SE \\
Sparse method        & Uniform \\
\hline\hline
\end{tabular}
\label{GAPparamstab}
\end{table}

\section{Bayesian Optimisation of GAP hyperparameters}\label{BOforGAP}

Crucial to the performance of a MLP is the careful selection of hyperparameters. While these are often chosen based on physical intuition, we instead adopt an iterative data-driven approach using Bayesian optimisation. This strategy is particularly effective for GAP, since the training cost with low-dimensional descriptors is relatively modest (especially in the parallelised implementation \cite{klawohn2023massively}).\\

As a first step, we determined the cutoff distances and number of sparse points for the two-body and three-body descriptors through convergence testing on the energy RMSE of a test set. The remaining hyperparameters ($\theta_{2b}, \delta_{2b}, \theta_{3b}, \delta_{3b}$) were then optimised using Bayesian optimisation with BOtorch\cite{balandat2020botorch}. The choice of hyperparameters for the EAM descriptor was informed by the converged two-body and three-body results. The four-dimensional hyperparameter space was initially explored using 10 uniformly sampled points. A Gaussian process prior with a radial basis function kernel was then used to model the objective function:

\begin{equation}
\mathbf{x}^* = \arg \min_{\mathbf{x} \in \mathcal{X}} 
\sqrt{ \frac{1}{N} \sum_{i=1}^{N} 
\left\| \mathbf{F}_i^{\text{ML}}(\mathbf{x}) - \mathbf{F}_i^{\text{DFT}} \right\|^2 }
\end{equation}

Here, $\mathbf{F}$ is a force component, $N$ is the number of samples, $\mathbf{x}$ is the vector of hyperparameters, $\mathcal{X}$ is the search space for hyperparameters, and $\mathbf{x}^*$ is the vector of optimised hyperparameters. The forces are derived from a validation set containing defects, perfect crystals and strained crystals of YBCO$_6$ and YBCO$_7$. The Gaussian process defines a mean and standard deviation for every sampled hyperparameter set. Well behaved hyperparameter priors from \cite{hvarfner2024vanilla} are utilised, and the log expected improvement acquisition function is used to decide which point to sample in the next iteration. The results of the optimisation are shown in supplementary figure \ref{BO}. Note that the force errors are high as the hyperparameter tuning was performed on an early version of the potential, and the validation set contained many high energy configurations.\\

\begin{figure*}[ht!]
    \centering
    \includegraphics[width=0.7\linewidth]{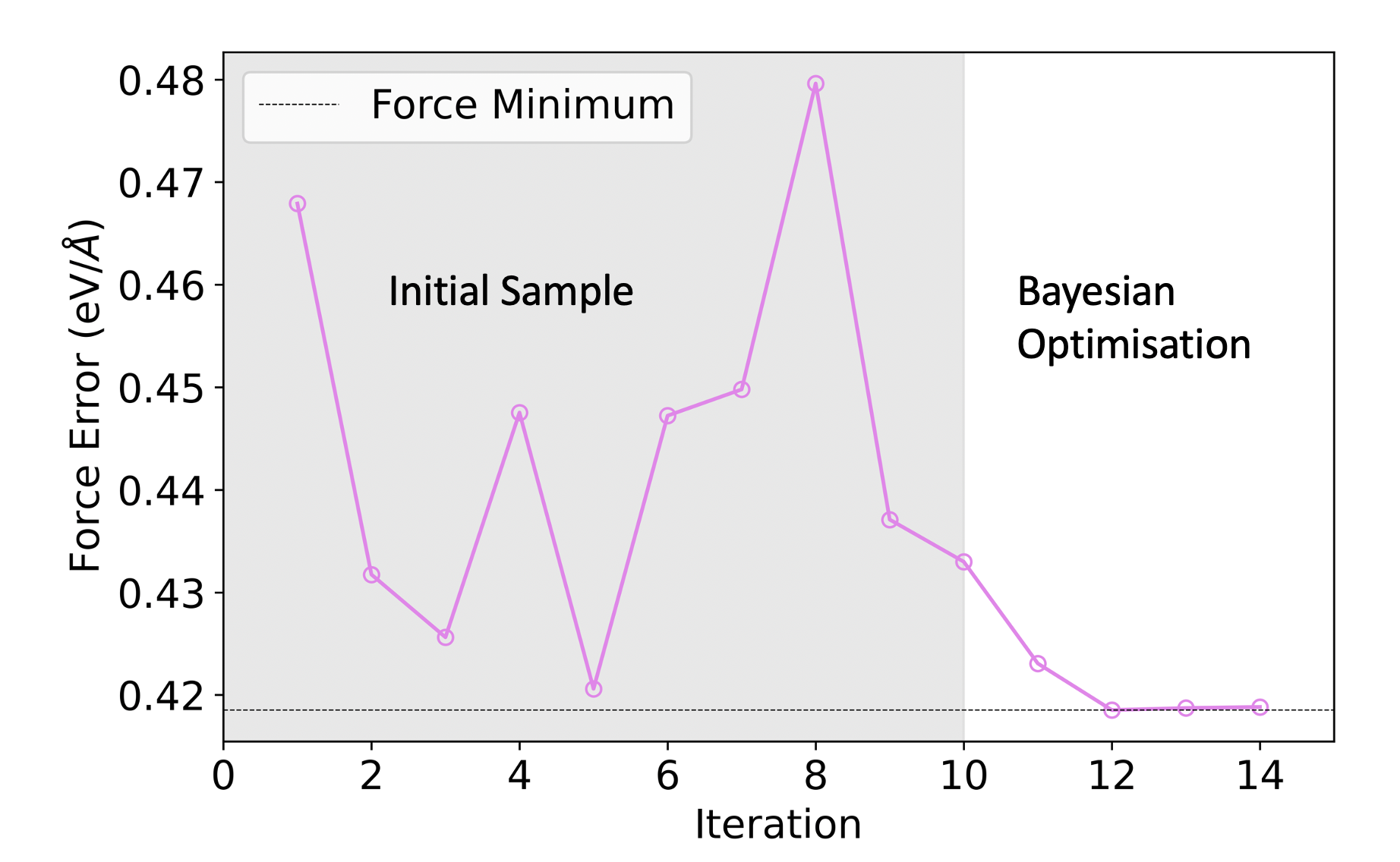}
    \caption{Force error versus iteration number for Bayesian optimisation of model parameters. The first 10 iterations comprise the initial samples, whereas the final 4 are selected via Bayesian optimisation.}
    \label{BO}
\end{figure*}

\section{Parameters of the tabGAP potential}

The tabGAP was tabulated using converged grids for all descriptors. Our tests indicated that 5000 grid points for the two-body and EAM descriptors were sufficient for a force error below 1 meV/{\AA}. An 80$\times$80$\times$80 grid was used for the three-body descriptor to again achieve a force error below 1 meV/{\AA}. The convergence testing for the three-body grid of the tabGAP is shown in supplementary figure \ref{convtabgap}.  
\begin{figure*}[ht!]
    \centering
    \includegraphics[width=0.7\linewidth]{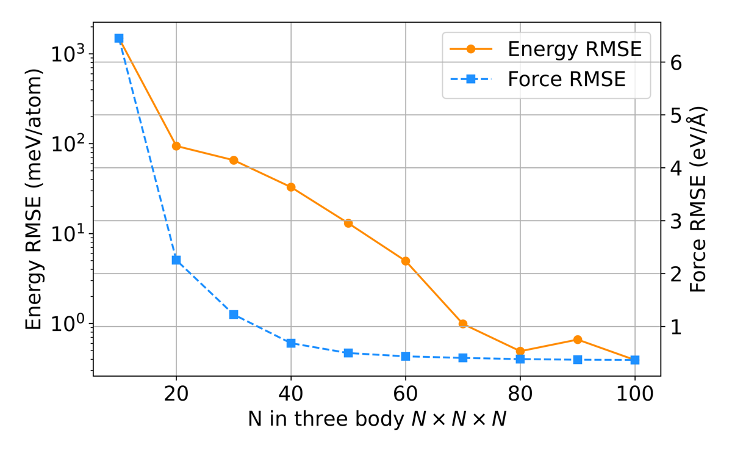}
    \caption{Convergence of force and energy RMSE for tabulation of GAP versus increasing grid density.}
    \label{convtabgap}
\end{figure*}

\section{DFT Lattice Parameters}

To validate the accuracy of our DFT simulations, the lattice parameters of the relaxed structures of each crystal subsystem are compared to those determined experimentally. The results are shown in supplementary table \ref{tablatticeparams}. In all cases the agreement is fair.  

\begin{table}[h!]
\centering
\caption{DFT and experimental lattice parameters of elemental solids, binary oxides, and the O$_2$ dimer. DFT values are given first, followed by experimental values.}
\label{tablatticeparams}
\begin{tabular}{lccc}
\hline\hline
Compound & $a$ (\AA) & $b$ (\AA) & $c$ (\AA) \\
\hline
Ba               & 5.13 / 5.03 \cite{evers1992lattice}      & 5.13 / 5.03      & 5.13 / 5.03 \\
Cu               & 3.65 / 3.61 \cite{straumanis1969lattice} & 3.65 / 3.61      & 3.65 / 3.61 \\
Y                & 3.65 / 3.65 \cite{spedding1956crystal}   & 3.65 / 3.65      & 5.76 / 5.73 \\
Cu$_2$O          & 4.34 / 4.27 \cite{brandt2014structural}  & 4.34 / 4.27      & 4.34 / 4.27 \\
BaO              & 5.64 / 5.54 \cite{zollweg1955x}          & 5.64 / 5.54      & 5.64 / 5.54 \\
Y$_2$O$_3$       & 10.68 / 10.60 \cite{gaboriaud2016disorder} & 10.68 / 10.60   & 10.68 / 10.60 \\
YBa$_2$Cu$_3$O$_7$ & 3.87 / 3.82 \cite{williams1988joint}    & 3.95 / 3.86      & 11.92 / 11.68 \\
O$_2$            & 1.24 / 1.21 \cite{chieh2007bond}         & N/A              & N/A \\
\hline\hline
\end{tabular}
\end{table}

\section{Phonon Spectra}

The phonon spectra predicted by all MLPs are shown in supplementary figure \ref{fig:phonon_comparison}. They are compared to VASP DFT data from \cite{Gambino2025}. Phonons are calculated using the small displacement method. All MLPs predict YBCO$_{7}$ to be dynamically stable. We note a large discrepancy for all potentials at the $S$ point. This was analysed by Gambino \textit{et al.} \cite{Gambino2025}, who's ACE potential also struggled to match this phonon band. The explanation was that numerical noise on the small energy scales involved here made the mode difficult to capture, especially due to the dense K-point mesh used in the phonon simulations. This was required to retrieve results that reflected a dynamically stable structure, as the fermi surface of YBCO is highly complex. Given the lack of such extensive k-space sampling in the training database of the potential, it is unlikely that all bands will be captured. If one is interested in more precise phonon calculations, far more accurate DFT training data would need to be utilised.  

\begin{figure*}[ht!]
    \centering
    \includegraphics[width=0.95\textwidth]{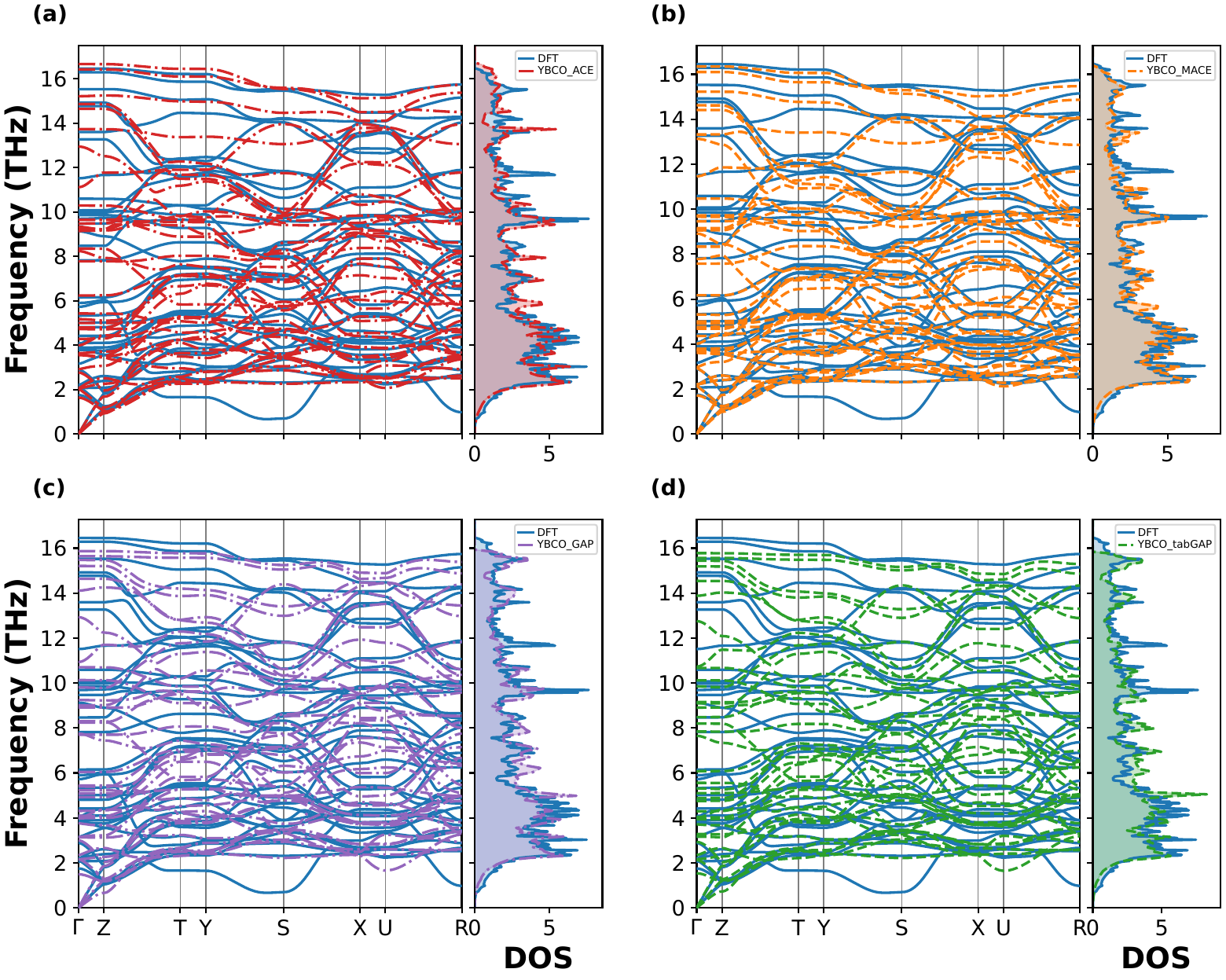}
    \caption{Comparison of phonon spectra of YBCO vs DFT obtained from (a) YBCO\_ACE, (b) YBCO\_MACE, (c) YBCO\_GAP and (d) YBCO\_tabGAP.}
    \label{fig:phonon_comparison}
\end{figure*}

\section{Chemical Potentials}

The calculated chemical potentials for each element in YBCO are shown in supplementary table \ref{chempot} (as described in the methodology section, no spin polarisation is employed). For Ba, Y and Cu, the chemical potential was calculated from the pure metal crystal, whereas for O it was calculated from the dimer. 

\begin{table}[h]
\centering
\caption{Chemical potentials from MACE, ACE, GAP, tabGAP, and DFT.}
\label{chempot}
\begin{tabular}{lccccc}
\hline\hline
Element & MACE (eV) & ACE (eV) & GAP (eV) & tabGAP (eV) & DFT (eV) \\
\hline
Ba & -693.45 & -693.45 & -693.47 & -693.47 & -693.45 \\
Y  & -1042.44 & -1042.67 & -1042.65 & -1042.65 & -1042.66 \\
Cu & -1309.52 & -1309.51 & -1309.51 & -1309.51 & -1309.51 \\
O  & -433.62 & -433.67 & -431.57 & -431.57 & -433.99 \\
\hline\hline
\end{tabular}
\end{table}

\end{document}